\documentclass[useAMS,usenatbib]{mn2e}

\usepackage{graphicx}
\usepackage{amsmath}
\usepackage{amssymb}
\usepackage{upgreek}
\usepackage{multirow}

\newcommand{\ion}[2]{\mbox{#1\,\textsc{#2}}}

\newcommand{\varampl}[0]{a}
\newcommand{\varslop}[0]{b}
\newcommand{\varpoly}[0]{c}
\newcommand{\varbusy}[0]{B_{0}}
\newcommand{\varbgen}[0]{B_{1}}
\newcommand{\varbsmp}[0]{B_{2}}

\renewcommand{\vec}[1]{\boldsymbol{#1}}
\newcommand{\matr}[1]{\mathbf{#1}}

\newcommand{\spc}[0]{\phantom{0}}
\newcommand{\changed}[1]{#1}

\title[The busy function]{The busy function: a new analytic function for describing the integrated 21-cm spectral profile of galaxies}
\author[T.~Westmeier, R.~Jurek, D.~Obreschkow, B.~S.~Koribalski and L.~Staveley-Smith]{T.~Westmeier$^{1}$\thanks{E-mail:
tobias.westmeier@uwa.edu.au}, R.~Jurek$^{2}$\thanks{Co-first author; e-mail:
russell.jurek@gmail.com}, D.~Obreschkow$^{1}$, B.~S.~Koribalski$^{2}$ and L.~Staveley-Smith$^{1}$\\
$^{1}$ICRAR, M468, The University of Western Australia, 35~Stirling Highway, Crawley~WA~6009, Australia\\
$^{2}$Australia Telescope National Facility, CSIRO Astronomy and Space Science, PO~Box~76, Epping~NSW~1710, Australia}

\begin{document}
    \date{Accepted 1988 December 15. Received 1988 December 14; in original form 1988 October 11}
    \pagerange{\pageref{firstpage}--\pageref{lastpage}} \pubyear{2002}
    \maketitle
    \label{firstpage}
    
    \begin{abstract}
      Accurate parametrization of galaxies detected in the 21-cm \ion{H}{i}~emission is of fundamental importance to the measurement of commonly used indicators of galaxy evolution, including the Tully--Fisher relation and the \ion{H}{i}~mass function. Here, we propose a new analytic function, named the `busy function', that can be used to accurately describe the characteristic double-horn \ion{H}{i}~profile of many galaxies. The busy function is a continuous, differentiable function that consists of only two basic functions, the error function, $\mathrm{erf}(x)$, and a polynomial, $|x|^{n}$, of degree $n \ge 2$. We present the basic properties of the busy function and illustrate its great flexibility in fitting a wide range of \ion{H}{i}~profiles from the Gaussian profiles of dwarf galaxies to the broad, asymmetric double-€horn profiles of spiral galaxies.
      
      Applications of the busy function include the accurate and efficient parametrization of observed \ion{H}{i}~spectra of galaxies and the construction of spectral templates for simulations and matched filtering algorithms. We demonstrate the busy function's power by automatically fitting it to the \ion{H}{i}~spectra of 1000~galaxies from the HIPASS Bright Galaxy Catalog, using our own C/C++ implementation, and comparing the resulting parameters with the catalogued ones. We also present two methods for determining the uncertainties of observational parameters derived from the fit.

    \end{abstract}
    
    
    \section{Introduction}
    
    Observations of the 21-cm emission line of neutral hydrogen provide measurements of several important parameters of galaxies, including their redshift, mass, and rotational velocity, as well as evolutionary indicators such as the Tully--Fisher relation \citep{Tully1977} and the \ion{H}{i}~mass function \citep{Zwaan1997}. While high-resolution \ion{H}{i}~maps have been obtained for several hundred nearby galaxies using radio interferometers, the vast majority of catalogued \ion{H}{i}~properties of galaxies has been extracted from integrated spectra obtained with single-dish telescopes. Over the next decade, \ion{H}{i}~surveys with the Square Kilometre Array (SKA; \citealt{Dewdney2009}) and some of its precursor and pathfinder instruments, such as ASKAP \citep{deBoer2009} and Apertif \citep{Oosterloo2009}, will probe larger volumes of space to much greater depth than ever before. As a result, the number of \ion{H}{i}-detected galaxies is expected to rise from currently $\ga 30\,000$ (HYPERLEDA; \citealt{Paturel2003}) to more than half a million galaxies predicted for the planned all-sky surveys WALLABY and WNSHS \citep{Duffy2012}. Yet, even in interferometric surveys like WALLABY, $95$~per cent of all expected detections will be less than three beams across (assuming a beam size of $30~\mathrm{arcsec}$) and hence only marginally resolved, highlighting the importance of accurate parametrization methods based on the integrated \ion{H}{i}~spectrum.
    
    \begin{figure*}
        \centering
        \includegraphics[width=0.9\linewidth]{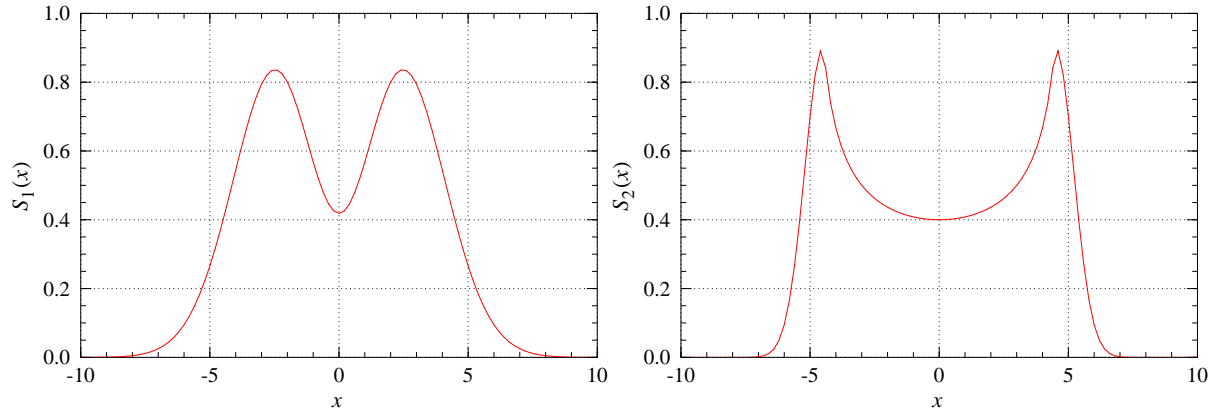}
        \caption{Previous attempts to describe the integrated \ion{H}{i}~spectra of galaxies. Left: Example of a spectral profile consisting of two Hermite functions as introduced by \citet{Saintonge2007} and defined in Eq.~\ref{eqn_saintonge}. The parameters used in creating this example are $a_{0} = 1$, $\varpoly = 0.3$, and $\sigma = 2$. Right: Example of the profile shape introduced by \citet{Obreschkow2009a,Obreschkow2009b} as specified in Eq.~\ref{eqn_obreschkow}. The parameters in this case are $k_{1} = 4.5$, $k_{2} = 1$, $k_{3} = 0.9$, $k_{4} = 25$, and $k_{5} = 2$.}
        \label{fig_saintonge}
    \end{figure*}
    
    Integrated \ion{H}{i}~line profiles encode much physical information. For example, (i)~the frequency centroid of the \ion{H}{i}~line measures the cosmological redshift plus peculiar motion of the galaxy, (ii)~the integral of the \ion{H}{i}~line provides a direct measure of the total \ion{H}{i}~mass \citep{Roberts1962}, (iii)~the line width traces the projected circular velocity of the galaxy and hence its dynamical mass \citep{Casertano1980}, and (iv)~the small-scale structure of the line profile encodes information on turbulent motion and warps \citep{Sancisi1976}. Furthermore, the shape of an \ion{H}{i}~line sensibly depends on disc asymmetries \citep{Baldwin1980}, extra-€planar gas \citep{Swaters1997,Heald2011}, tidal tails, and companions. Finally, observational settings, such as spectral resolution and noise level, also affect observed \ion{H}{i}~lines. The efficient extraction of all this information from thousands of noisy \ion{H}{i}~spectra requires a quick and accurate parametrization method.
    
    Here we present the `busy function', $\varbusy$, a heuristic, analytic function suitable for the parametrization of galaxy \ion{H}{i}~spectra, as well as two modifications, $\varbgen$ and $\varbsmp$. We explore their characteristics and potential applications. Furthermore, we apply the function to the 1000~\ion{H}{i}~spectra from the HIPASS Bright Galaxy Catalog \citep{Koribalski2004} and compare the results with the published \ion{H}{i}~properties of the galaxies. Possible applications of the busy function include the extraction of galaxy properties from spectral profile fitting, the construction of profile shapes for matched-filtering techniques in automated source finding, and the possibility of analytically describing complex spectral profiles of galaxies using only a few basic parameters. \changed{This paper focuses on the application of spectral fitting for the purpose of accurate galaxy parametrisation, an aspect driven by the need for automated parametrisation of large samples of galaxies in future} \ion{H}{i} \changed{surveys with the SKA and its precursors.}
    
    The paper is structured as follows: In Section~\ref{sect_previouswork} we discuss previous approaches to fitting the spectral profiles of galaxies. In Section~\ref{sect_busyfunction} we introduce the busy function and its basic properties. Two variations of the busy function are described in Section~\ref{sect_modifications}, while examples of fitting the function to the integrated spectra of real galaxies are presented in Section~\ref{sect_examples}. In Section~\ref{sect_hipassbgc} we demonstrate the power of the busy function by automatically fitting it to the 1000~spectra of the HIPASS Bright Galaxy Catalog and comparing our results with the catalogued parameters of the galaxies. Section~\ref{sect_conclusions} summarises our results and conclusions. In the Appendix we present some of the more technical aspects of the paper. Appendix~\ref{app_analytical} discusses some of the analytic solutions of the busy function. In Appendix~\ref{app_uncertainties} we present and discuss methods for determining the uncertainties of parameters derived from busy function fits, while Appendix~\ref{app_gaussian} explores the relationship between the busy function and a Gaussian. Finally, in Appendix~\ref{app_software} we present the details of the C/C++ implementation of the busy function fitting algorithm developed for this paper.

    \begin{figure*}
        \centering
        \includegraphics[width=0.9\linewidth]{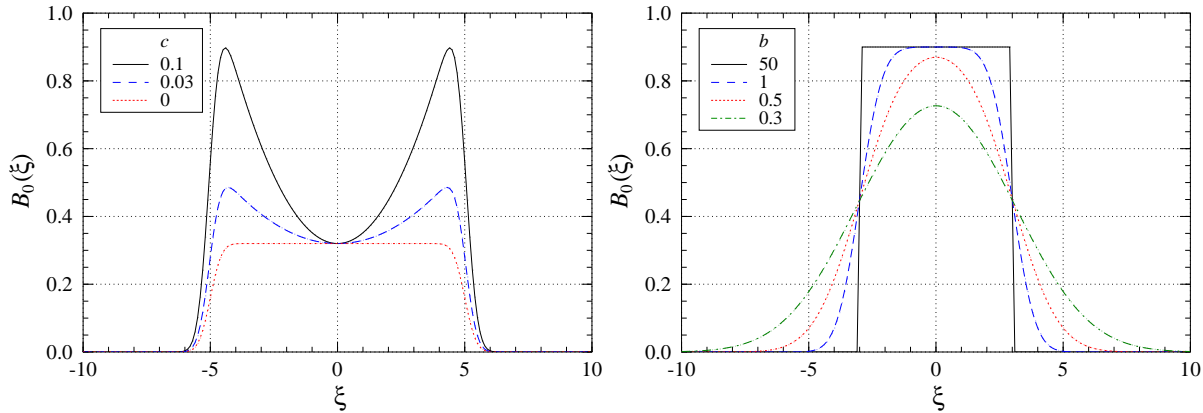}
        \caption{The busy function, $\varbusy$, for different values of $\varpoly{}$ (left) and $\varslop{}$ (right). The left-hand panel shows the situation for $w = 5$, $\varampl{} = 0.32$, $\varslop{} = 2$, and varying $\varpoly{}$. The right-hand panel depicts $w = 3$, $\varampl{} = 0.9$, $\varpoly{} = 0$, and varying $\varslop{}$.}
        \label{fig_busyfunction}
    \end{figure*}

    \section{Previous work}
    \label{sect_previouswork}
    
     Previous attempts \changed{to describe the double-horn profiles of galaxy spectra} usually involved Gaussian components. For the purpose of \ion{H}{i}~source finding in the extragalactic ALFALFA survey \citep{Giovanelli2005} on the Arecibo telescope, \citet{Saintonge2007} employed a matched-filtering technique using analytic template spectra. As a template she chose a linear combination of two Hermite functions of the form
    \begin{equation}
        S_{1}(x) = a_{0} \Psi_{0}(x, \sigma) + \varpoly \Psi_{2}(x, \sigma) , \label{eqn_saintonge}
    \end{equation}
    where $\Psi_{0}(x, \sigma)$ and $\Psi_{2}(x, \sigma)$ are the first two symmetric Hermite functions, and $a_{0}$ and $\varpoly$ are free parameters. An example of this profile shape is shown in the left-hand panel of Fig.~\ref{eqn_saintonge}.
    
    While the profile bears some resemblance to the general double-peaked profiles seen in the \ion{H}{i}~spectra of galaxies, it differs from the true shape of most galaxy spectra. Firstly, the overall shape of the profile is close to Gaussian, and hence the flanks are not steep enough to reflect the sharp rise generally seen in the spectra of observed galaxies, in particular large spiral galaxies. Secondly, the central trough does not resemble the generally wide and flat troughs actually seen in many disc galaxies. In addition, the peaks of the profile are much smoother than the sharp peaks often seen in observed \ion{H}{i}~spectra. However, a great advantage of the approach by \citet{Saintonge2007} is the small number of free parameters required to describe the profile, \changed{making it ideal for fulfilling its original purpose of serving as a template for matched filtering.}
     
     A more realistic profile was developed by \citet{Obreschkow2009a,Obreschkow2009b} for the description of \ion{H}{i}~discs generated by semi-analytic models. They used a combination of two different functions to model the outer flanks and the central trough of the profile separately, namely
    \begin{equation}
        S_{2}(x) =
        \begin{cases}
            k_{3} \exp \! \left( \! -\frac{[|x| - k_{1}]^{2}}{k_{2}} \right) & \text{for}~|x| \ge \Delta x_{\rm p} / 2 , \\
	    \frac{k_{5}}{\sqrt{k_{4} - x^{2}}}                               & \text{for}~|x| < \Delta x_{\rm p} / 2 , \\
        \end{cases}
        \label{eqn_obreschkow}
    \end{equation}
    where $\Delta x_{\rm p}$ is the separation between the two peaks. This heuristic profile shape was motivated by the desire to closely match the line shape derived from the modelling of an edge-on, rotating gas disc using simple analytic functions. An example of this profile is shown in the right-hand panel of Fig.~\ref{fig_saintonge}.
    
    While providing a fairly accurate description of the sharp and narrow peaks and broad troughs of typical galaxy spectra, there are a few disadvantages with this approach. Firstly, the profile gets broken up into two separate functions that would have to be fitted separately. Secondly, precise adjustment of the parameters is required to avoid creating large discontinuities at the boundary between the two functions. Another complication arises from the function used to create the central trough of the profile, which is undefined for $x^{2} \ge k_{4}$; this may pose a problem for fitting algorithms. In contrast to the profile used by \citet{Saintonge2007}, the profile of \citet{Obreschkow2009a,Obreschkow2009b} can reproduce the steep flanks and sharp peaks of observed galaxy spectra by decoupling the width of the Gaussian component from the overall profile width.

    \section{The busy function}
    \label{sect_busyfunction}
    
    \subsection{Definition}
    
    In order to improve on previous attempts to \changed{describe the double-horn profile of galaxies}, we looked for a function that would allow us to describe the steep flanks often seen in the spectra of galaxies while also recovering the characteristic trough and sharp, narrow peaks of the spectrum. In addition, we require the function to be continuous and differentiable for the purpose of least-squares fitting, \changed{for which calculation of the partial derivatives with respect to the function's free parameters, $\partial f(x, \vec{p}) / \partial p_{i}$, is required.} Such a function can indeed be constructed in a relatively simple fashion by multiplying two error functions and a parabola,
    \begin{align}
        \varbusy(\xi) &= \frac{\varampl{}}{4}
                         \times (\mathrm{erf}[\varslop{} \{ w + \xi \} ] + 1) \nonumber \\
                      &  \times (\mathrm{erf}[\varslop{} \{ w - \xi \} ] + 1)
                         \times \left( \varpoly{} \, \xi^{2} + 1 \right) \! ,
        \label{eqn_busyfunction}
    \end{align}
    with $\xi \equiv x - x_{0}$. Here, the variable $x$ represents the spectral axis of the data, e.g.\ frequency or radial velocity. The two error functions, $\mathrm{erf}(x)$, generate the profile flanks, while multiplication with a parabola produces the central trough of the profile. Given its characteristics and versatility, we chose to call this function the `busy function'. Examples of the busy function with different parameter values are shown in Fig.~\ref{fig_busyfunction}. \changed{In this section we will first discuss the fundamental properties of the basic form of the busy function as specified in Eq.~\ref{eqn_busyfunction} before introducing a more general, asymmetric form of the function in Section~\ref{sect_modifications}.}
    
    \subsection{Free parameters}
    
    The busy function is characterised by five free parameters: the centroid of the profile, $x_{0}$; the half-width of the profile, $w$; the total amplitude scaling factor, $\varampl{}$; and two additional parameters, $\varslop{}$ and $\varpoly{}$. The parameter $\varslop{}$ controls the steepness of the two error functions constituting the flanks of the spectrum. In the case of $\varslop{} \rightarrow \infty$ the flanks will become infinitely steep, whereas for $\varslop{} \rightarrow 0$ the slope of the flanks will approach zero. The parameter $\varpoly{}$ controls the emphasis of the parabola and hence the amplitude of the central trough. Values of $\varpoly{} > 0$ imply increasing amplitudes of the trough, whereas for $\varpoly{} = 0$ the trough will disappear altogether. Negative values of $\varampl$, $\varslop{}$, $\varpoly{}$, and $w$, while mathematically allowed, are not physically meaningful in the case of \ion{H}{i}~emission lines, but may be useful in other situations not considered here, such as absorption lines.
    
    \begin{figure*}
        \centering
        \includegraphics[width=0.9\linewidth]{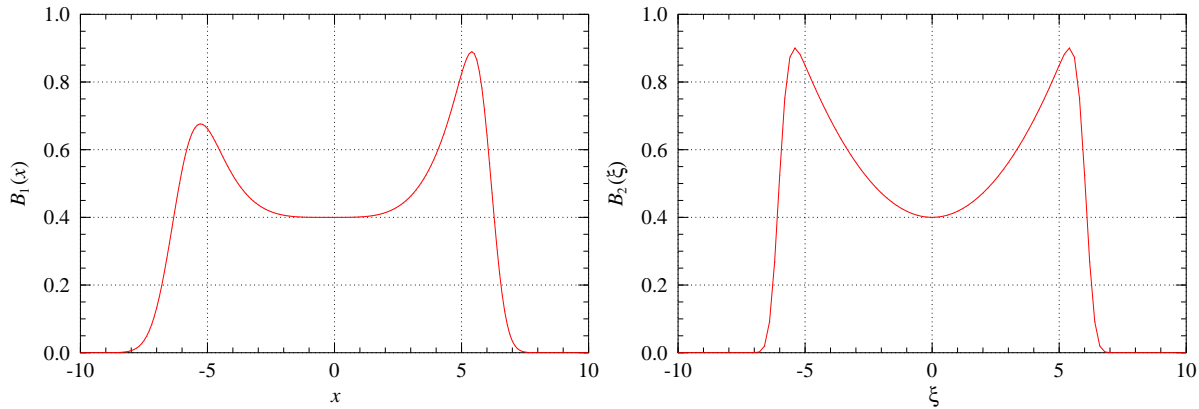}
        \caption{Left: Example of the asymmetric, generalised busy function, $\varbgen(x)$, for $\varslop{}_{1} = 1$, $\varslop{}_{2} = 1.5$, $\varpoly{} = 0.0015$, $x_{\rm e} = 0$, $x_{\rm p} = -0.2$, and $n = 4$,. Right: Example of the simplified busy function, $\varbsmp(\xi)$, for $\varslop{} = 0.2$ and $\varpoly{} = 0.045$. In both cases, $\varampl{} = 0.4$ and $w = 6$.}
        \label{fig_modifiedbusyfunction}
    \end{figure*}
    
    \subsection{General properties}
    
    Some of the analytic solutions to the busy function are elaborated and presented in Appendix~\ref{app_analytical}. For profiles with $\varslop{} w \gg 1$ (i.e.~flat-topped or double-horn profiles), the value at the centre of the profile is simply given by $\varbusy(0) = \varampl{}$. The profile's half-width, $w$, is equal to half the separation of the two error functions, and hence the full width at half maximum of the profile is equal to $2 w$ in the case of $\varpoly{} = 0$ and $\varslop{} w \gg 1$.
    
    An advantage of the busy function, and the motivation for its name, is its versatility when it comes to fitting spectral profiles of different shape. Examples of the busy function mimicking double-peaked profiles of different shape are presented in the left-hand panel of Fig.~\ref{fig_busyfunction}. By carefully choosing appropriate values for $\varslop{}$, $\varpoly{}$, and $w$, almost any shape of (symmetric) double-peaked profile can be reproduced by the function. By using error functions to represent the flanks of the spectrum, we can reproduce the characteristic, steep rise of the spectral profile typically observed in the integrated \ion{H}{i}~spectra of disc galaxies.
    
    The flexibility of the busy function goes well beyond the fitting of double-peaked profiles, as is illustrated in the right-hand panel of Fig.~\ref{fig_busyfunction}. Here, the parameter $\varpoly{}$ was set to zero to entirely remove the parabolic component and, in combination with different values of $\varslop{}$, produce profiles of varying shape ranging from a steep top-hat function (with large $\varslop{}$) to a Gaussian function with gradual slopes on both sides (using smaller values of $\varslop{}$). In fact, with the right choice of parameters, the product of two error functions will take almost the exact shape of a Gaussian function (see Appendix~\ref{app_gaussian}). Hence, the busy function is capable of fitting both Gaussian and double-peaked line profiles.
    
    \section{Modifications of the busy function}
    \label{sect_modifications}
    
    \subsection{Generalised busy function}
    \label{sect_gbf}
    
    In its original form of Eq.~\ref{eqn_busyfunction}, the busy function is symmetric, and the shape of its central trough is determined by the parabolic component. However, at the expense of additional free parameters, the busy function can be generalised to describe a wider range of spectral profiles, e.g.\ profiles that are not symmetric or have a differently shaped trough. In a more general form, the busy function can be written as
    \begin{align}
        \varbgen(x) &= \frac{\varampl{}}{4}
                \times (\mathrm{erf}[\varslop{}_{1} \{ w + x - x_{\rm e} \} ] + 1) \nonumber \\
             &  \times (\mathrm{erf}[\varslop{}_{2} \{ w - x + x_{\rm e} \} ] + 1)
                \times \left( \varpoly{} \, |x - x_{\rm p}|^{n} + 1 \right) \! .
        \label{eqn_busyfunction2}
    \end{align}
    The number of free parameters in this case has increased from five to eight, including separate slopes, $\varslop_{1}$ and $\varslop_{2}$, for the two error functions, separate offsets, $x_{\rm e}$ and $x_{\rm p}$, for the error functions and the polynomial, and a variable degree, $n$, of the polynomial. The properties of the generalised busy function are similar to those of the original busy function and discussed in Appendix~\ref{app_analytical}.
    
    An example of an asymmetric, generalised busy function with a fourth-degree polynomial trough ($n = 4$) is depicted in the left-hand panel of Fig.~\ref{fig_modifiedbusyfunction}. The significantly broader trough more closely resembles those typically found in the spectra of many spiral galaxies.
    
    \subsection{Simplified busy function}
    
    A simplification of the busy function can be achieved by replacing the two error functions, $\mathrm{erf}(\xi)$, with just a single error function with the argument, $\xi$, squared. While the shape and properties of the resulting function are very similar to the original busy function, the expressions for the simplified busy function, and in particular its derivatives, are less complex and significantly shorter:
    \begin{equation}
        \varbsmp(\xi) = \frac{\varampl{}}{2}
        \times \left(\mathrm{erf} \! \left[ \varslop{} \left\{ w^{2} - \xi^{2} \right\} \right] + 1 \right)
        \times \left( \varpoly{} \, \xi^{2} + 1 \right) .
        \label{eqn_busyfunction3}
    \end{equation}
    As before, we define $\xi \equiv x - x_{0}$ for simplicity. This function, as depicted in the right-hand panel of Fig.~\ref{fig_modifiedbusyfunction}, is very similar to the original busy function as specified in Eq.~\ref{eqn_busyfunction}. Like $\varbusy$, the simplified busy function can be generalised by introducing an independent $x_{0}$ for the parabola or by replacing the parabolic trough with a different function, e.g.\ a fourth-degree polynomial. The slopes of the two flanks, however, will always be the same.
    
    Some of the analytic solutions to the simplified busy function are discussed in Appendix~\ref{app_analytical}. Unlike $\varbusy$, there is no combination of parameters for which the profile would closely resemble a Gaussian function. In the best approximation, $\varbsmp$ will appear more compact than a Gaussian, having steeper flanks and a slightly smaller amplitude.
    
    \begin{figure*}
        \centering
        \includegraphics[width=0.45\linewidth]{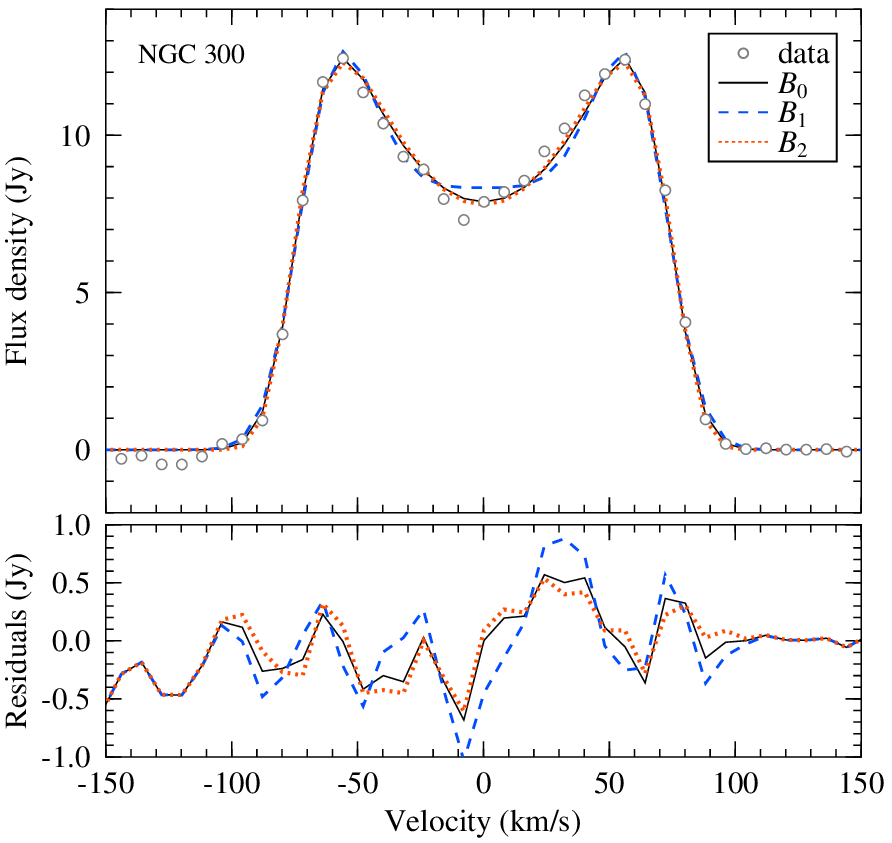}
        \includegraphics[width=0.45\linewidth]{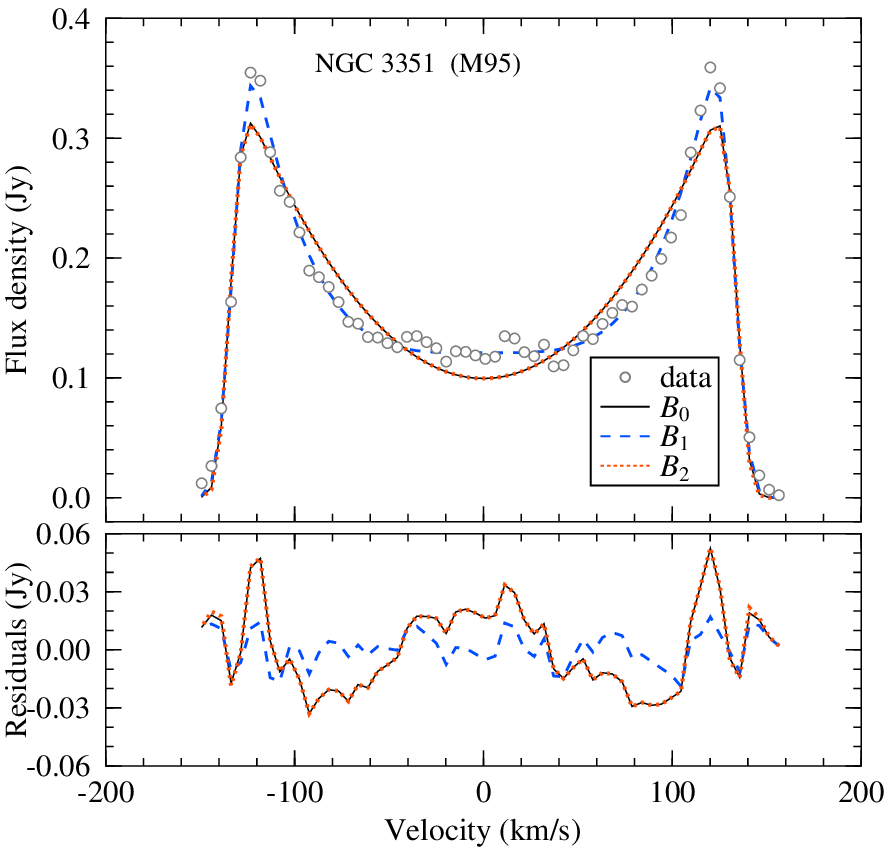}
        \caption{Top panels: Fits of symmetric versions of the busy functions $\varbusy$, $\varbgen$, and $\varbsmp$ to the integrated \ion{H}{i}~spectra of NGC~300 (left) and NGC~3351 (right). In the case of $\varbgen$ we assumed a polynomial trough of degree $n = 4$. The bottom panels show the residuals between the data and the fits.}
        \label{fig_ngc300}
    \end{figure*}

    \section{Examples}
    \label{sect_examples}
    
    In this section we present a few examples of fitting the busy function to the \ion{H}{i}~spectra of observed galaxies to illustrate its usefulness and flexibility.
    
    \subsection{Symmetric profiles}
    
    In Fig.~\ref{fig_ngc300} we show integrated \ion{H}{i}~spectra of the two spiral galaxies NGC~300 \citep{Westmeier2011} and NGC~3351\,/\,M95 \citep{Walter2008}. All three versions of the busy function, $\varbusy$, $\varbgen$, and $\varbsmp$, have been fitted to the spectra using a $\chi^{2}$~minimization algorithm. In the case of the generalised busy function, $\varbgen$, we assumed symmetry ($\varslop_{1} = \varslop_{2}$, $x_{\rm e} = x_{\rm p}$), but used a fourth-degree polynomial ($n = 4$) to generate a broader trough.
    
    NGC~300 is an example of a medium-sized spiral galaxy with a symmetric \ion{H}{i}~spectrum and a relatively sharp, almost V-shaped trough that is equally well described by either the original busy function, $\varbusy$, or the simplified busy function, $\varbsmp$, both of which use a second-degree polynomial to describe the trough. In contrast to these, the much wider fourth-degree polynomial trough chosen for the generalised busy function, $\varbgen$, does not describe the appearance of NGC~300 very well. This is obvious from the residuals between the data and model, which are significantly larger ($\sigma_{\rm rms} = 0.46~\mathrm{Jy}$) for $\varbgen$ as compared to the other two fits ($\sigma_{\rm rms} \approx 0.32~\mathrm{Jy}$).
    
    Note that the residuals in this and all following examples are due to real structures and asymmetries in the galaxies themselves and much larger than expected from the baseline noise of the integrated spectra alone.
    
   NGC~3351 (M95) is an example of a galaxy with a broad trough. In this case, the generalised busy function, $\varbgen$, with its fourth-degree polynomial provides a much better fit ($\sigma_{\rm rms} = 8.8~\mathrm{mJy}$) to the integrated \ion{H}{i}~spectrum than the busy functions $\varbusy$ and $\varbsmp$ ($\sigma_{\rm rms} \approx 21.2~\mathrm{mJy}$).
    
    \begin{figure*}
        \centering
        \includegraphics[width=0.45\linewidth]{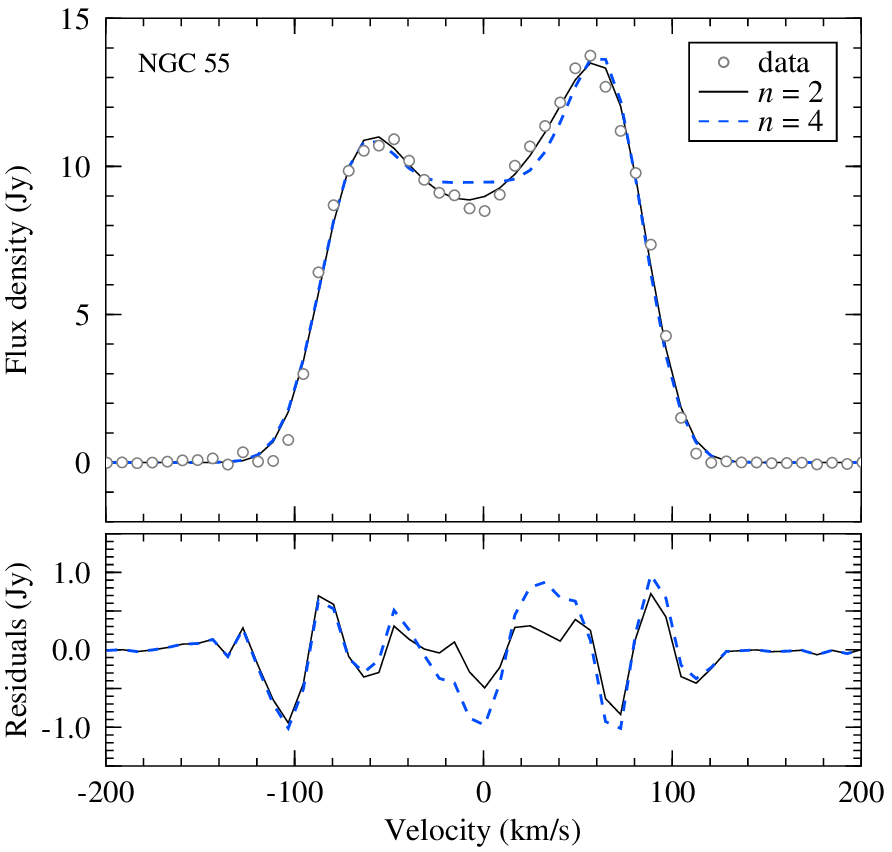}
        \includegraphics[width=0.45\linewidth]{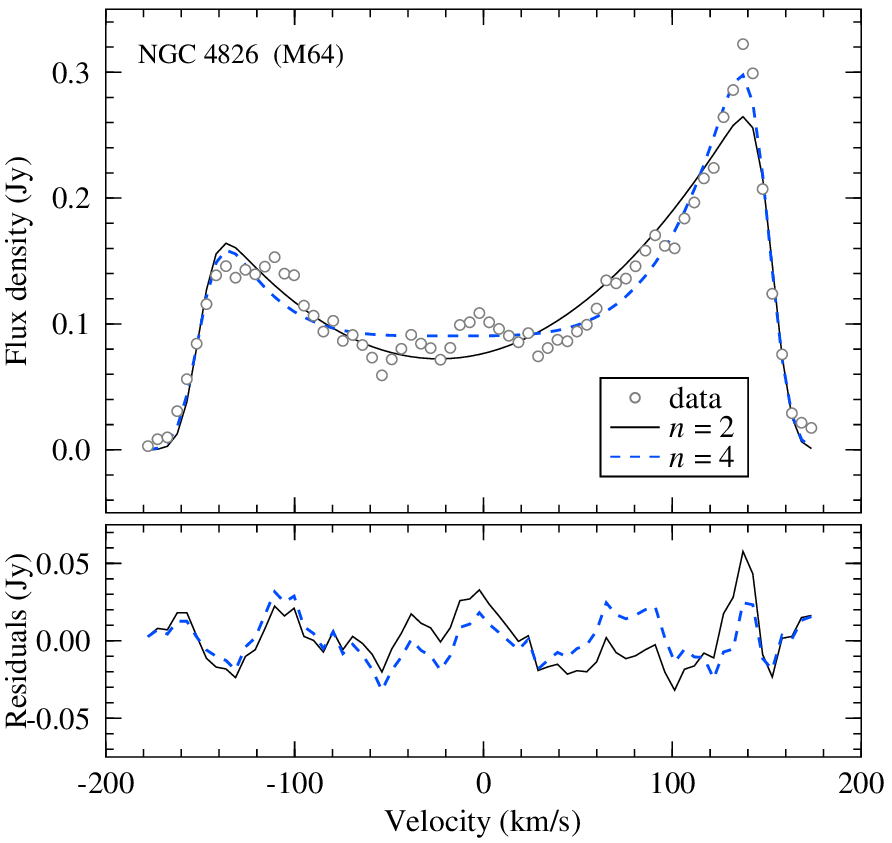}
        \caption{Top panels: Fits of two asymmetric versions ($x_{\rm e} \ne x_{\rm p}$, but $\varslop_{1} = \varslop_{2}$) of the generalised busy function, $\varbgen$, with different polynomial troughs of degree $n = 2$ and~$4$ to the integrated \ion{H}{i}~spectra of NGC~55 (left) and NGC~4826 (right). The bottom panels show the residuals between the data and the fits.}
        \label{fig_ngc55}
    \end{figure*}
  
    \subsection{Asymmetric profiles}
    
    As discussed in Section~\ref{sect_modifications}, the busy function can be easily generalised to fit asymmetric profiles by simply introducing a separate offset for the polynomial component describing the trough. Two examples of asymmetric busy functions fitted to the integrated \ion{H}{i}~spectra of the two galaxies NGC~55 \citep{Westmeier2013} and NGC~4826 (M64; \citealt{Walter2008}) are presented in Fig.~\ref{fig_ngc55}. Here, we fitted two versions of the generalised busy function, $\varbgen$, to the spectrum, this time including separate offsets, $x_{\rm e}$ and $x_{\rm p}$, for the error functions and the polynomial trough. This allows the trough to shift with respect to the flanks of the spectrum, producing an asymmetric profile with peaks of different height. The two functions use polynomial degrees of $n = 2$ and $4$, respectively, while adopting a single slope for the two flanks ($\varslop_{1} = \varslop_{2}$).
    
    The integrated spectrum of NGC~55 resembles that of NGC~300 with the exception of a noticeable asymmetry that is well fitted by the generalised busy function. As in the case of NGC~300, the parabolic trough ($n = 2$) fits better ($\sigma_{\rm rms} = 0.41~\mathrm{Jy}$) than the fourth-degree polynomial ($\sigma_{\rm rms} = 0.62~\mathrm{Jy}$). Within their uncertainties, the offsets significantly differ between the overall spectrum described by the error functions, $x_{\rm e} = 119.4 \pm 0.4~\mathrm{km \, s}^{-1}$, and the parabolic trough, $x_{\rm p} = 110.6 \pm 1.3~\mathrm{km \, s}^{-1}$, thereby quantitatively confirming the intrinsic asymmetry of the spectrum.
    
    The spectrum of NGC~4826 is clearly asymmetric, too. As before, both versions of the generalised busy function provide a good fit to the spectrum, although the fourth-degree polynomial is capable of fitting the broad trough and sharp peaks in the spectrum much better ($\sigma_{\rm rms} = 14.0~\mathrm{mJy}$) than the parabolic trough ($\sigma_{\rm rms} = 17.3~\mathrm{mJy}$). Again, there is a significant difference between the location of the overall spectrum, $x_{\rm e} = 409.3 \pm 0.7~\mathrm{km \, s}^{-1}$, and that of the fourth-degree polynomial trough, $x_{\rm p} = 391.5 \pm 2.4~\mathrm{km \, s}^{-1}$, confirming the intrinsic asymmetry of the spectrum.

   \begin{table*}
  	\centering
	\caption{Comparison of catalogued and calculated HIPASS BGC properties for direct extraction of parameters from the spectrum as well as parametrization based on busy function fitting. For each parameter and signal-to-noise ratio the table lists two different components of the comparison: the result of a linear regression with the catalogued parameter as the independent variable (best-fit slope and intercept; see Fig.~\ref{fig_bgcfit}) and the fraction of parameters within a certain percentage of the original, catalogued values.}
	\label{table_compObs}
	Integrated flux ($F_{\rm int}$) \\
	\begin{tabular}{lrrrrrr}
	\hline
	 & \multicolumn{2}{c}{original} & \multicolumn{2}{c}{$S/N = 5$} & \multicolumn{2}{c}{$S/N = 3$} \\
	 & direct & BF fit & direct & BF fit & direct & BF fit \\
	\hline
	best-fit slope                                     &  $1.02$ &  $1.03$ & $0.96$ &  $1.05$ & $0.92$ & $1.07$ \\
	best-fit intercept ($\mathrm{Jy \, km \, s}^{-1}$) & $-1.08$ & $-1.35$ & $2.01$ & $-1.02$ & $4.49$ & $0.14$ \\
	within  5\% of cat. (\%) \hspace{1cm}              &  $89.8$ &  $83.3$ & $35.8$ &  $37.1$ & $22.3$ & $23.2$ \\
	within 10\% of cat. (\%)                           &  $96.9$ &  $95.5$ & $65.6$ &  $66.9$ & $42.5$ & $43.2$ \\
	within 25\% of cat. (\%)                           &  $98.9$ &  $99.1$ & $93.2$ &  $91.4$ & $77.4$ & $76.4$ \\
	\hline
	\end{tabular} \\
	Peak flux density ($F_{\rm peak}$) \\
	\begin{tabular}{lrrrrrr}
	\hline
	 & \multicolumn{2}{c}{original} & \multicolumn{2}{c}{$S/N = 5$} & \multicolumn{2}{c}{$S/N = 3$} \\
	 & direct & BF fit & direct & BF fit & direct & BF fit \\
	\hline
	best-fit slope                        & $1.00$ & $0.99$ &  $1.14$ &  $1.04$ & $1.27$ & $1.09$ \\
	best-fit intercept (Jy)               & $0.00$ & $0.00$ & $-0.01$ & $-0.01$ & $0.01$ & $0.00$ \\
	within  5\% of cat. (\%) \hspace{1cm} & $99.0$ & $68.6$ &  $21.9$ &  $28.3$ &  $5.4$ & $18.1$ \\
	within 10\% of cat. (\%)              & $99.2$ & $88.0$ &  $43.3$ &  $54.5$ & $14.7$ & $32.8$ \\
	within 25\% of cat. (\%)              & $99.7$ & $99.1$ &  $86.3$ &  $91.9$ & $47.1$ & $72.2$ \\
	\hline
	\end{tabular} \\
	Line width ($w_{50}$) \\
	\begin{tabular}{lrrrrrr}
	\hline
	 & \multicolumn{2}{c}{original} & \multicolumn{2}{c}{$S/N = 5$} & \multicolumn{2}{c}{$S/N = 3$} \\
	 & direct & BF fit & direct & BF fit & direct & BF fit \\
	\hline
	best-fit slope                               & $0.98$ & $0.99$ & $0.94$ & $0.95$ & $0.86$ & $0.87$ \\
	best-fit intercept ($\mathrm{km \, s}^{-1}$) & $0.64$ & $1.47$ & $12.9$ & $6.39$ & $60.2$ & $21.7$ \\
	within  5\% of cat. (\%) \hspace{1cm}        & $99.4$ & $86.6$ & $46.4$ & $50.3$ & $25.1$ & $29.0$ \\
	within 10\% of cat. (\%)                     & $99.4$ & $94.0$ & $69.4$ & $72.2$ & $42.4$ & $47.8$ \\
	within 25\% of cat. (\%)                     & $99.6$ & $97.7$ & $86.0$ & $89.3$ & $63.2$ & $72.1$ \\
	\hline
	\end{tabular} \\
	Line width ($w_{20}$) \\
	\begin{tabular}{lrrrrrr}
	\hline
	 & \multicolumn{2}{c}{original} & \multicolumn{2}{c}{$S/N = 5$} & \multicolumn{2}{c}{$S/N = 3$} \\
	 & direct & BF fit & direct & BF fit & direct & BF fit \\
	\hline
	best-fit slope                               & $0.96$ & $0.96$ & $0.90$ & $0.87$ & $0.89$ & $0.80$ \\
	best-fit intercept ($\mathrm{km \, s}^{-1}$) &  $8.7$ &  $5.6$ & $62.2$ & $39.4$ & $94.9$ & $83.7$ \\
	within  5\% of cat. (\%) \hspace{1cm}        & $94.0$ & $86.0$ & $34.6$ & $47.7$ & $20.7$ & $26.5$ \\
	within 10\% of cat. (\%)                     & $94.6$ & $93.2$ & $53.0$ & $68.1$ & $36.8$ & $45.4$ \\
	within 25\% of cat. (\%)                     & $96.7$ & $97.9$ & $72.2$ & $82.0$ & $56.7$ & $65.0$ \\
	\hline
	\end{tabular} \\
  \end{table*}

  \section{Practical Application}
  \label{sect_hipassbgc}
  
  In this section we present the practical application of busy function fitting to the integrated spectra of the HIPASS Bright Galaxy Catalog (BGC) sources \citep{Koribalski2004}. We present both the methodology and the results. Most of the sources in the BGC are individual galaxies with a unique optical counterpart, although 68~sources are flagged as confused, 44 identified as pairs, and 11 identified as compact groups. 91~detections do not have an optical identification, mostly as a result of Galactic foreground extinction.
  
  \subsection{Method}
  
  We developed our own software to fit the busy function to all 1000 galaxies in the BGC. The technical details of our implementation are presented in Appendix~\ref{app_software}. Our software attempts to fit six variants of the busy function with a varying number of free parameters, using the Levenberg-Marquardt algorithm (LMA; \citealt{Levenberg1944,Marquardt1963}). The software then selects the best of these fits based on the Akaike Information Criterion \citep{Akaike1974} in an attempt to determine the optimal number of free parameters needed to describe the data. A key feature of our method is the use of multiple (several thousand), short duration (tens of iterations) LMA attempts for each model, each attempt starting from a randomly chosen position in parameter space. This implementation is available in the form of C and C++ libraries and as a python module (see Appendix~\ref{app_software} for details). The advantages of our implementation are:
  \begin{enumerate}
    \item Model fits do not suffer from parameter discretisation.
    \item The method ensures rapid and efficient exploration of multi-dimensional parameter spaces.
    \item A covariance matrix is produced for the model fit, thus providing parameter uncertainties as well as correlations.
    \item The likelihood of finding the best global fit is higher.
    \item The code takes advantage of multi-core CPUs, distributed systems, and GPUs.
    \item The use of model variants allows us to set model components to exactly zero (true zeroing). This is otherwise impossible in practice because of the limited numerical precision of computers.
    \item The algorithm takes into account the uncertainties of individual data points, resulting in data points with large uncertainties to be effectively excluded from fitting.
    \item The software can easily be made to fit any analytic model or function.
  \end{enumerate}
  \changed{Our implementation of the busy function fitting algorithm can be obtained from a dedicated website\footnote{http://code.google.com/p/busy-function-fitting/} as a C~library, a C++~template library, and a Python module.}
  
  \subsection{Results}
  
  We successfully fitted the busy function to all 1000~HIPASS BGC sources. We accomplished this by using an iterative strategy. In each iteration we fitted the busy function to all BGC sources that were not successfully fitted in the previous iteration, each time using different random LMA starting positions in parameter space. \changed{The details of this procedure are explained in Appendix~\ref{app_software}. Success or failure of a fit was assessed by checking whether the reduced $\chi^{2}$ of the fit was reasonable. In addition, the quality and accuracy of the fit was visually confirmed.} The success rate was about $85$~per cent for each iteration, and in the fourth and fifth iterations we were only processing two and a single BGC source, respectively. The success of this iterative approach confirms that the failures in any iteration are purely a result of poor LMA starting positions, because they are randomly chosen in each iteration. This also confirms our expectation that we could improve the success rate of each iteration, by increasing the number of LMA starting positions (at an increased computational cost). \changed{Alternatively, initial estimates of the free parameters, e.g.~from a preceding source finding run, can be used instead of random starting positions in parameter space, thereby avoiding the need for multiple iterations altogether.}
  
  \begin{figure*}
      \centering
      \includegraphics[width=0.85\linewidth]{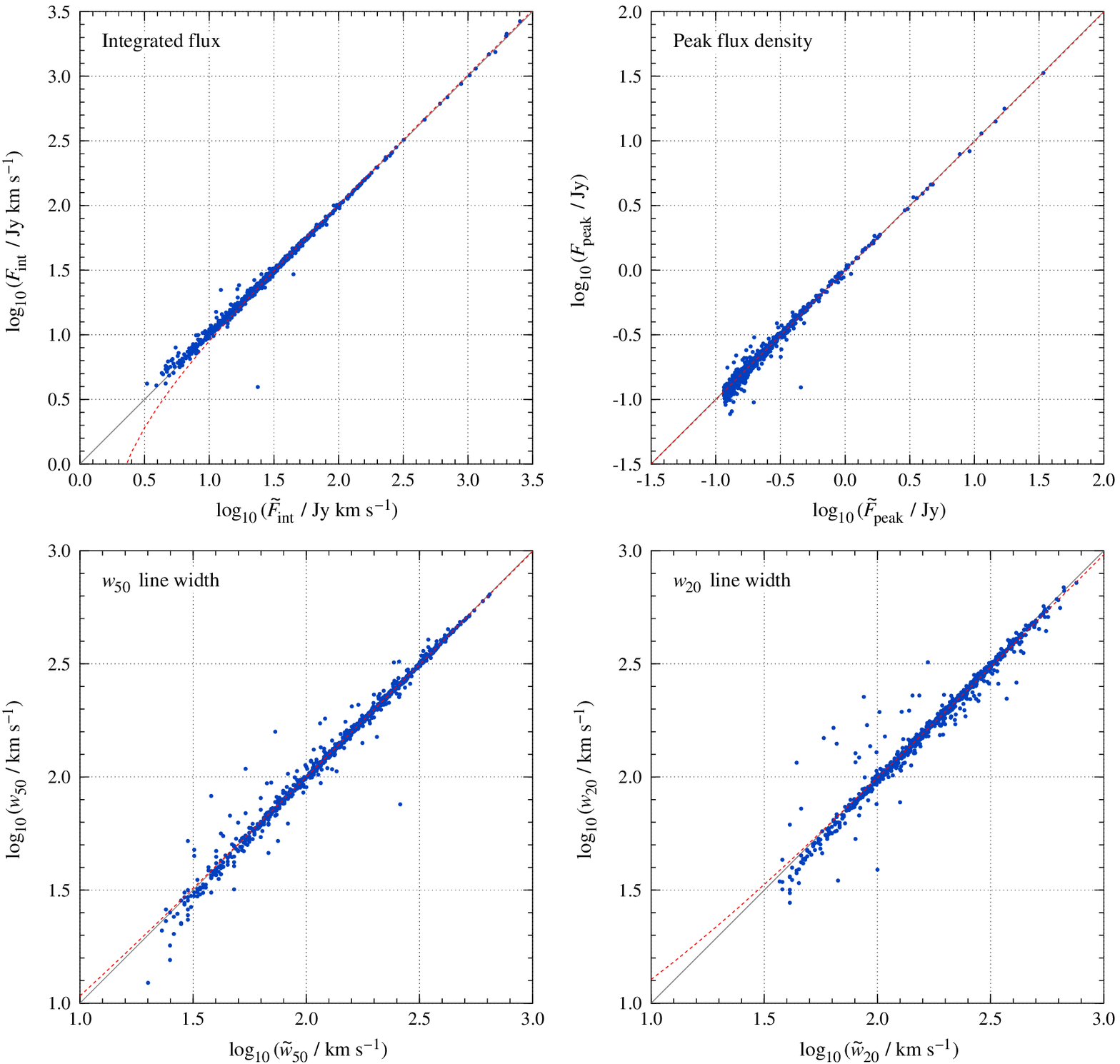}
      \caption{Comparison of the integrated flux, peak flux density, and $w_{50}$/$w_{20}$~line widths of the HIPASS BGC galaxies as derived from the busy function fit (ordinate) with the original values listed in the BGC (abscissa; parameters marked with tilde). The solid, grey line represents the identity, while the dashed, red line is the result of a linear regression carried out in linear space (see Table~\ref{table_compObs}).}
      \label{fig_bgcfit}
  \end{figure*}
  
  To test the performance of the busy function in the case of spectra with lower signal-to-noise ratio, we injected additional noise into the HIPASS BGC spectra to generate spectra with a peak signal-to-noise ratio of three and five. We then re-fitted the busy function, using the same iterative procedure based on visual inspection. The success rate for both sets of noisier BGC spectra was approximately $90$~per cent. After six iterations we were able to fit a `sensible' busy function (as qualitatively judged by us) to every noisy BGC spectrum.
  
  In Table~\ref{table_compObs} and Fig.~\ref{fig_bgcfit} we compare the catalogued observational properties of BGC sources against those derived from the busy function fits, allowing us to infer the quality of the fits. \changed{The catalogued parameters simply serve as a reference and are not necessarily unbiased or more accurate than our fitting results.} We compare the integrated \ion{H}{i}~flux ($F_{\rm int}$), peak \ion{H}{i}~flux density ($F_{\rm peak}$), and the spectral line widths at $20$ and $50$~per cent of the peak flux density ($w_{20}$ and $w_{50}$, respectively). The comparison was carried out using both a busy function sampled at the same spectral resolution as the data (SD) and a high-definition (HD) busy function sampled at 100~times the data's spectral resolution. It turns out that there is no discernible benefit to using the HD busy function over the SD busy function, and hence only the SD~results are listed in Table~\ref{table_compObs}. We believe that this is a result of the catalogued properties being derived directly from the data. We also re-calculated the observational properties directly from the data, \changed{following the same approach as described in section~3.3 of the BGC paper \citep{Koribalski2004}}. These `direct' properties \changed{are used as a sanity check and} are calculated in the same manner as the catalogued properties, but using the channel range within which the fitted busy function is $\ge 1$~per cent of its peak value. Any differences between the direct and catalogued properties should solely be due to differences in the channel range used in the measurement.
  
  Fig.~\ref{fig_bgcfit} presents a comparison of our parametrization of the HIPASS BGC spectra (without additional noise) with the original measurement of each parameter in the BGC. We accurately recover the fluxes and line widths of the galaxies from the fitted busy function, with differences of usually well under $10$~per cent relative to the original, catalogued parameters. The results illustrate the great accuracy with which galaxy parameters can be derived from busy function fits. The few outliers seen in Fig.~\ref{fig_bgcfit} are mostly due to artefacts in the data, including interference and variations in the spectral baseline. \changed{Such effects can have a strong impact on the accuracy of the parametrisation and will potentially affect any parametrisation method. Some of the outliers in Fig.~\ref{fig_bgcfit} in particular could be due to the additional baseline subtraction carried out for the BGC catalog, but not for the busy function fitting, potentially resulting in discrepancies for individual galaxies affected by baseline artefacts.}
  
  In Table~\ref{table_compObs} the calculated and catalogued HIPASS BGC observational properties are compared in two ways. Firstly, a linear regression is carried out to test if a 1:1~correlation exists. The result of the linear regression is shown as the dashed, red curve in Fig.~\ref{fig_bgcfit} (note that for a non-vanishing intercept the linear fit appears curved in logarithmic space). Secondly, we measure the fraction of calculated values that lie within $5$, $10$ and $25$~per cent of the catalogued values. In doing so, we implicitly ignore the uncertainties in the original BGC parameters, although these would contribute to the measured differences as well. The uncertainties published by \citet{Koribalski2004} are in the order of 10~per cent for $F_{\rm peak}$, $w_{50}$, and $w_{20}$, and about 15~per cent for $F_{\rm int}$.
   
  Two main conclusions can be drawn from Table~\ref{table_compObs}. Firstly, as the signal-to-noise ratio decreases the busy function fits recover the catalogued properties more reliably than direct measurement of the properties does. This is most evident for the peak flux density and, to a lesser degree, for the velocity widths. \changed{The reason for the busy function to perform better is that, at low signal-to-noise ratio, the direct parametrisation method may be strongly affected by noise, whereas the fit is much less affected by individual noise peaks.} There is little advantage, however, when measuring the integrated flux. It should be noted that this is to be expected with Gaussian noise, provided that the \changed{chosen} channel range contains all of the signal, \changed{because the emission is integrated over the entire source}. Secondly, despite the differences in the catalogued and calculated properties, we find that the majority of properties derived from the busy function fits are within $5$, $10$, and $25$~per cent of the catalogued properties at signal-to-noise ratios of $\gtrsim 10$, $5$, and $3$, respectively.
  
  Our analysis demonstrates that the catalogued HIPASS BGC properties can be reliably recovered in a fully automated approach for the majority of sources across a range of signal-to-noise ratios. Potential applications include the parametrization of a large sample of galaxies as well as the construction of realistic \ion{H}{i}~profiles for simulations such as the S$^3$--SAX simulation \citep{Obreschkow2009b}. Additionally, storing the fitted busy function parameters in addition to the full galaxy spectra will be of particular benefit to large future \ion{H}{i}~surveys such as WALLABY and DINGO \citep{Duffy2012,Meyer2009}, allowing them to include a simplified representation of every integrated \ion{H}{i}~spectrum in their source catalogue using a maximum of just eight parameter values.
  
  Another great advantage of galaxy parametrization through the method of busy function fitting is the possibility to determine the statistical uncertainties of derived observational parameters from the uncertainties of the fit. This will enable a full error analysis based on the fitting results alone and without the need to modify the input data for that purpose. In Appendix~\ref{app_uncertainties} we present the detailed description and analysis of two methods to determine uncertainties of observational parameters from the covariance matrix provided by least-squares fitting algorithms.
  
   \section{Conclusions}
   \label{sect_conclusions}
   
   We present a continuous and differentiable analytic function, called the busy function, $\varbusy$, and two modifications, $\varbgen$ and $\varbsmp$, designed to describe the typical double-horn profile commonly observed in the integrated \ion{H}{i}~spectra of spiral galaxies. With a set of five to eight free parameters, the busy function accurately describes a wide range of spectral profiles of galaxies, including symmetric and asymmetric double-horn profiles, simple Gaussian profiles, and flat-topped profiles with steep flanks.
   
   The most promising application of the busy function, \changed{and the main focus of this paper}, is the possibility to automatically fit the integrated \ion{H}{i}~spectra of a large sample of galaxies. This will allow common observational parameters of galaxies, such as the line width or the integrated flux, to be measured with great accuracy and in an automated fashion. In addition, a simple functional representation of each galaxy's integrated spectrum with a maximum of just eight parameters can be stored in a source catalogue. Another potential application of the busy function, \changed{although not investigated in this paper}, is the generation of a sample of analytic mock profiles of galaxy spectra to be used as templates, either for the purpose of modelling or in matched filtering algorithms of source finding pipelines.
   
   A great advantage of parametrizing galaxies by fitting a busy function to the integrated spectrum is the possibility to determine statistical uncertainties of the derived observational parameters. In Appendix~\ref{app_uncertainties} we present two methods that allow the uncertainties of observational parameters to be determined without the need to modify the input data. This will enable a proper error analysis even in situations where classical Monte-Carlo or bootstrap methods cannot be applied, e.g.~when a single spectrum of low signal-to-noise ratio needs to be parametrized.
   
   In order to test the suitability of the busy function for automated spectral-line fitting, we implemented a Levenberg-Marquardt algorithm in C/C++ (Appendix~\ref{app_software}) to fit busy functions to the 1000~galaxies of the HIPASS BGC. Our results demonstrate that it is possible to fit the busy function to a large number of galaxy spectra in a fully automated way without any human intervention. A comparison of several measured galaxy parameters (integrated flux, peak flux density, and $w_{20}$ and $w_{50}$ line widths) with those listed in the HIPASS Bright Galaxy Catalog reveals that we accurately recover the observational parameters of the galaxies from the fit.
   
   For the original spectra almost all of our measured parameters lie within $25$~per cent of the catalogued ones. Even when reducing the peak signal-to-noise ratio of the spectra to~5 and~3, that fraction still remains at about $90$ and $70$~per cent, respectively. In addition, our measurement based on the busy function fit is often more accurate than the direct parameter measurement carried out on the spectrum. This result illustrates another great strength of the busy function: parametrization based on fitting a busy function to the spectrum is less strongly affected by the noise in individual channels and thus produces more accurate results than any direct measurement. As a consequence, the number of galaxies in an observational sample that can be successfully parametrized would potentially increase compared to conventional parametrization methods, thereby improving the accuracy of scientific studies such as the measurement of the Tully--Fisher relation.
   
   While originally designed to describe the integrated \ion{H}{i}~emission spectra of galaxies, the busy function's versatility will allow it to be used in many other areas, for instance in the fitting of \ion{H}{i}~absorption spectra (see \citealt{Allison2013} for an actual example), stacked \ion{H}{i}~spectra of galaxies (e.g.~\citealt{Fabello2011,Delhaize2013}), and integrated CO spectra of galaxies \changed{(e.g.~\citealt{Young2011,Saintonge2011,Tacconi2013})}.

   \section*{Acknowledgements}
    
   We thank the members of the \ion{H}{i}~source finding collaboration for stimulating discussions that led to the development of the busy function. The Australia Telescope is funded by the Commonwealth of Australia for operation as a National Facility managed by CSIRO. This work made use of THINGS, `The \ion{H}{i} Nearby Galaxy Survey' \citep{Walter2008}.

    \appendix
    
    \section{Analytic solutions of the busy function}
    \label{app_analytical}
    
    In this section we describe some of the analytic solutions for the evaluation of the three busy functions, $B_{n}$, with $n \in \{0, 1, 2\}$, as defined in Eq.~\ref{eqn_busyfunction}, \ref{eqn_busyfunction2}, and \ref{eqn_busyfunction3}. For simplicity, we will only consider symmetric functions here (i.e.\ $\varslop_{1} = \varslop_{2}$ and $x_{\rm e} = x_{\rm p}$ for $\varbgen$).
    
    It is straightforward to calculate the derivatives of the busy function with respect to $x$ or any of the free parameters, and we will refrain from presenting the rather lengthy analytic expressions here. By setting $\mathrm{d} B_{n} / \mathrm{d} x = 0$ we would in principle be able to calculate the positions of the extrema of $B_{n}$, but we have been unable to find a solution to this equation for any of the variants of the busy function, and there may not be an analytic solution other than the trivial solution of $x = x_{0}$ (i.e.\ $\xi = 0$). Hence, numerical methods will need to be used to determine the positions of the peaks of $B_{n}$ as well as the resulting galaxy parameters, such as peak flux density or profile width.
    
    For all three versions of the busy function we can calculate the value at the position of the trivial extremum, $\xi = 0$, which corresponds to the centre of the profile. The value at the centre of the original busy function, $\varbusy$, and the generalised busy function, $\varbgen$, is given by
    \begin{equation}
      \varbusy(0) = \varbgen(0) = \frac{\varampl}{4} \, (\mathrm{erf}[\varslop w] + 1)^{2} ,
    \end{equation}
    which will converge to $\varampl$ for $\varslop w \gg 1$ (i.e.~broad profiles with steep flanks). Note that this only applies to the symmetric version of $\varbgen$ where $\varslop_{1} = \varslop_{2}$ and $x_{\rm e} = x_{\rm p}$. The value at the centre of the simplified busy function, $\varbsmp$, is given by
    \begin{equation}
        \varbsmp(0) = \frac{\varampl}{2} \left( \mathrm{erf} \! \left[ \varslop w^{2} \right] + 1 \right) \! ,
    \end{equation}
    which will again converge to $\varampl$ for $\varslop w^{2} \gg 1$.
    
    While we cannot produce general solutions for the profile width, there is a simple solution for cases where the polynomial component disappears, i.e.~$\varpoly = 0$, and the flanks of the spectrum are sufficiently steep, i.e.~large values of $\varslop w$ (for $\varbusy$ and $\varbgen$) or $\varslop w^{2}$ (for $\varbsmp$). In this case of `boxy' spectra, the separation between the two error functions of $2 w$ is equal to the full width at half maximum (or $w_{50}$) of the profile. In other cases of double-horn profiles with steep flanks, although not exact, this solution may still provide a first-order estimate of $w_{50}$.

    \section{Relation between the busy function and the Gaussian function}
    \label{app_gaussian}
    
    \begin{figure}
        \centering
        \includegraphics[width=0.9\linewidth]{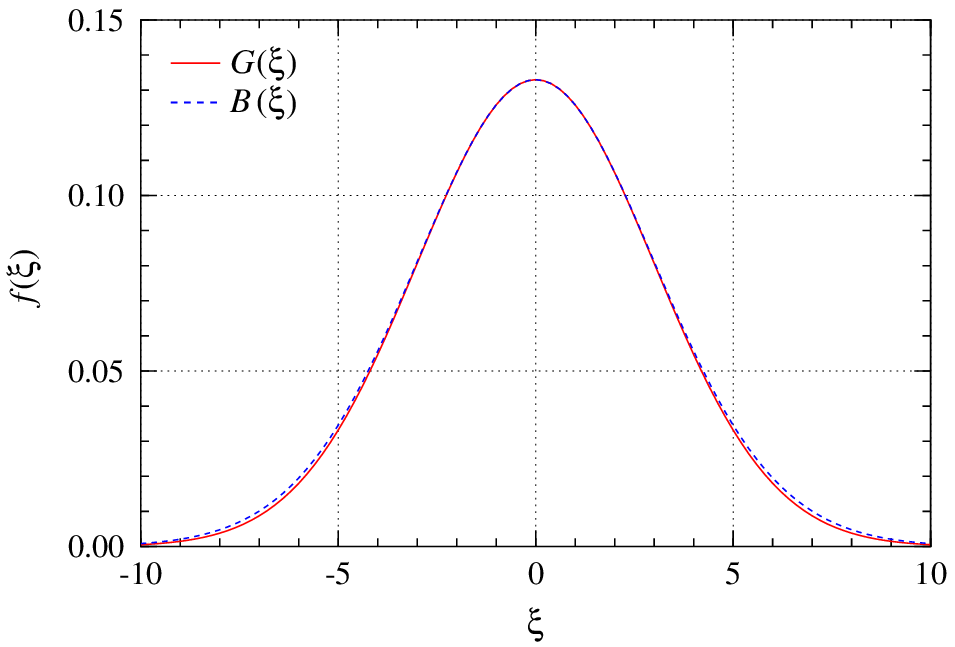}
        \caption{Comparison of the busy function, $B(\xi)$, with a Gaussian function, $G(\xi)$, of dispersion $\sigma = 3$. The parameters of the two functions are linked according to Eq.~\ref{eqn_a0} and~\ref{eqn_a1}.}
        \label{fig_bg}
    \end{figure}
    
    \changed{The Gaussian function is commonly used for the purpose of fitting spectral lines, including the} \ion{H}{i}~\changed{spectra of galaxies and gas clouds. In this section we demonstrate that the original busy function, $\varbusy$, without the polynomial trough (i.e.~$\varpoly = 0$) possesses the flexibility to closely approximate a Gaussian function and hence can describe the generally simple line profiles of dwarf galaxies and face-on galaxies just as well.} In order to achieve this, let us first define a normalised Gaussian function of the form
    \begin{equation}
      G(\xi) = \frac{1}{\sqrt{2 \uppi} \, \sigma} \exp \! \left( \! -\frac{\xi^{2}}{2 \sigma^2} \right)
    \end{equation}
    and a busy function similar to Eq.~\ref{eqn_busyfunction}, but in a simplified form, assuming a width of $w = 0$ and no polynomial component ($\varpoly = 0$), hence
    \begin{equation}
      B(\xi) = \varampl (\mathrm{erf}[\varslop \xi] + 1) (\mathrm{erf}[-\varslop \xi] + 1) . \label{eqn_busy}
    \end{equation}
    For simplicity, we define $\xi \equiv x - x_{0}$, as before. Both functions can be expanded into a Taylor series at the origin, $\xi = 0$:
    \begin{alignat}{9}
      G(\xi) & = \; & \frac{1}{\sqrt{2 \uppi} \, \sigma} \xi^{0} & \; - \; & \frac{1}{\sqrt{8 \uppi} \, \sigma^{3}} \xi^2 & \; + \; & \frac{1}{\sqrt{128 \uppi} \, \sigma^{5}} \xi^{4} & \; + \; & \ldots \label{eqn_taylor_g} \\
      B(\xi) & = \; & \varampl \xi^{0} & \; - \; & \frac{4 \varampl \varslop^{2}}{\uppi} \xi^{2} & \; + \; & \frac{8 \varampl \varslop^{4}}{3 \uppi} \xi^{4} & \; + \; & \ldots \label{eqn_taylor_b}
    \end{alignat}
    Apparently, the Taylor series of the busy function and the Gaussian function are very similar. Both only contain terms of even order, and a comparison of the respective coefficients in each order should thus allow us to derive approximate relations between the parameters of the busy function and the standard deviation, $\sigma$, of the Gaussian function.
    
    From a comparison of the zeroth-order and second-order terms of the two Taylor series in Eq.~\ref{eqn_taylor_g} and~\ref{eqn_taylor_b} we obtain the following solutions for the parameters $\varampl$ and $\varslop$ of the busy function as defined in Eq.~\ref{eqn_busy}:
    \begin{align}
      \varampl &= \frac{1}{\sqrt{2 \uppi} \, \sigma} , \label{eqn_a0} \\
      \varslop &= \frac{\sqrt{\uppi}}{\sqrt{8} \, \sigma} = \frac{\uppi}{2} \, \varampl . \label{eqn_a1}
    \end{align}
    The two parameters above describe a family of busy functions that closely resemble a Gaussian function of width $\sigma$, although they are approximations only.
    
    An example is shown in Fig.~\ref{fig_bg}, where the red, solid curve shows a Gaussian function, $G(\xi)$, with a width of $\sigma = 3$, while the blue, dashed curve shows the busy function, $B(\xi)$, with parameters $\varampl = 1 / \sqrt{18 \uppi}$ and $\varslop = \sqrt{\uppi / 72}$ according to Eq.~\ref{eqn_a0} and~\ref{eqn_a1}. Both functions match almost perfectly, in particular near the origin of $\xi = 0$. For larger values of $|\xi|$ the relative difference between $B(\xi)$ and $G(\xi)$ increases, but the absolute difference remains small across the entire domain of the two functions. The busy function's remarkable resemblance of a Gaussian function adds to its versatility when it comes to fitting the wide range of different \ion{H}{i}~profiles found in galaxies.

    \section{Implementation of a busy function fitting algorithm}
    \label{app_software}
    
  \begin{table*}
	\centering
	\caption{The six busy function variants used in our C/C++/Python implementation, their dimensionality/complexity and the associated $\chi^{2}$ penalty imposed by the Akaike Information Criterion.}
	\label{table_BFvars}
	\begin{tabular}{ccl}
	\hline
	No.\ of free & AIC $\chi^{2}$ & busy function variant \\
	parameters   & penalty      & \\
	\hline
	$4$ & $+\spc{}8$ & $(\alpha / 4) \times (1 + \mathrm{erf}[\beta_{\spc} \{ x - \gamma_{1} \} ]) \times (1 + \mathrm{erf}[\beta_{\spc} \{ \gamma_{2} - x \} ])$ \\
	$5$ &      $+10$ & $(\alpha / 4) \times (1 + \mathrm{erf}[\beta_{1}    \{ x - \gamma_{1} \} ]) \times (1 + \mathrm{erf}[\beta_{2}    \{ \gamma_{2} - x \} ])$ \\
	$5$ &      $+10$ & $(\alpha / 4) \times (1 + \mathrm{erf}[\beta_{\spc} \{ x - \gamma_{1} \} ]) \times (1 + \mathrm{erf}[\beta_{\spc} \{ \gamma_{2} - x \} ]) \times \left( 1 + \phi \, | x - [0.5 \{ \gamma_{1} + \gamma_{2} \} ] |^{4} \right)$ \\
	$6$ &      $+12$ & $(\alpha / 4) \times (1 + \mathrm{erf}[\beta_{\spc} \{ x - \gamma_{1} \} ]) \times (1 + \mathrm{erf}[\beta_{\spc} \{ \gamma_{2} - x \} ]) \times \left( 1 + \phi \, | x - [0.5 \{ \gamma_{1} + \gamma_{2} \} ] |^{n} \right)$ \\
	$7$ &      $+14$ & $(\alpha / 4) \times (1 + \mathrm{erf}[\beta_{\spc} \{ x - \gamma_{1} \} ]) \times (1 + \mathrm{erf}[\beta_{\spc} \{ \gamma_{2} - x \} ]) \times \left( 1 + \phi \, | x - \theta |^{n} \right)$ \\
	$8$ &      $+16$ & $(\alpha / 4) \times (1 + \mathrm{erf}[\beta_{1}    \{ x - \gamma_{1} \} ]) \times (1 + \mathrm{erf}[\beta_{2}    \{ \gamma_{2} - x \} ]) \times \left( 1 + \phi \, | x - \theta |^{n} \right)$ \\
	\hline
	\end{tabular}
  \end{table*}
    
  \begin{table*}
	\centering
	\caption{The variable mappings used to impose parameter range constraints during fitting. For a given user-specified range, the values $mid$ and $amp$ are the mid-point and distance from either end of the range to the mid-point, respectively.}
	\label{table_BFmaps}
	\begin{tabular}{llll}
	\hline
	Variable(s) & Mapping & Range & Reason \\
	\hline
	$\alpha$                             & $\exp(\alpha)$              & $> 0$             & Meaningful normalisation; avoids normalisation trade-offs. \\
	$\beta$,  $\beta_{1}$,  $\beta_{2}$  & $\exp(\beta)$               & $> 0$             & Prevents inversion of error function origins ($\gamma$, $\gamma_{1}$, $\gamma_{2}$). \\
	$\gamma$, $\gamma_{1}$, $\gamma_{2}$ & $mid + amp \, \sin(\gamma)$ & user-specified    & Avoids irrelevant solutions. \\
	$\phi$                               & $\exp(\phi)$                & $> 0$             & Meaningful normalisation; avoids normalisation trade-offs. \\ 
	$\theta$                             & $mid + amp \, \sin(\theta)$ & user-specified    & Avoids irrelevant solutions. \\
	$n$                                  & $5 + 3 \, \sin(n)$          & $2 \leq n \leq 8$ & Avoids unphysical solutions. \\
	\hline 
	\end{tabular}
  \end{table*}
   
  \begin{table*}
	\caption{The parameter lists used to generate LMA starting positions, using random selection with replacement.}
 	\label{table_LVMseeds}
	\begin{tabular}{ll}
	\hline
	Property & LMA seed values \\
	\hline
	model peak & $1/8$, $1/4$, $3/8$, $1/2$, $5/8$, $3/4$, $7/8$, $1$ $\times$ data maximum \\
	$\beta^{(1)}$, $\beta_{1}^{(1)}$, $\beta_{2}^{(1)}$ & $0.1$, $0.2$, $0.4$, $0.8$, $1.6$, $3.2$, $6.4$, $10$, $g/1.25$, $g/2.5$, $g/5$, $g/10$, $g/15$, $g/20$ \\
	$\gamma_1^{(2)}$ & $min$, $+r/8$, $+r/4$, $+3r/8$, $+r/2$, $+5r/8$, $+3r/4$, $+7r/8$, $max$ \\
	$\gamma_2^{(2,3)}$ & $min$, $+r/8$, $+r/4$, $+3r/8$, $+r/2$, $+5r/8$, $+3r/4$, $+7r/8$, $max$ provided $\gamma_{2} \geq \gamma_{1}$ \\
	$\phi$ for $\alpha = 10^{-3}$ & $10^{-9}$, $10^{-8}$, $10^{-7}$, $10^{-6}$, $10^{-5}$, $10^{-4}$, $10^{-3}$, $0.01$, $0.1$, $1$, $10$, $100$, $10^{3}$, $10^{4}$ \\
	$\theta$ & $\gamma_{1}$, $(\gamma_{2} - \gamma_{1})/6$, $(\gamma_{2} - \gamma_{1})/3$, $(\gamma_{2} - \gamma_{1})/2$, $2(\gamma_{2} - \gamma_{1})/3$, $5(\gamma_{2} - \gamma_{1})/6$, $\gamma_{2}$ \\
	$n$ & $2$, $4$, $6$ \\
	\hline
	\end{tabular} \\
	Notes: (1)~The constant $g = 0.747806$ is used to calculate $\beta$ values corresponding to Gaussian roll-offs of the form, $\beta = g / \sigma$.
	       (2)~$min$ and $max$ are the user-specified roll-off range; $+r$ denotes $min$ plus the additional quantity, with $r \equiv max - min$.
	       (3)~$\gamma_{1}$ must be chosen first so that it can be used to limit the number of possible $\gamma_{2}$ values.
  \end{table*}
  
    We implemented a busy function fitting program using C++ and the C libraries \textsc{cfitsio} and \textsc{cpgplot}. Our fitting program is based on a combination of the Levenberg-Marquardt algorithm (LMA; \citealt{Levenberg1944,Marquardt1963}) and the Akaike Information Criterion (AIC; \citealt{Akaike1974}). We use the LMA to carry out a $\chi^{2}$ minimization for six variants of the busy function as listed in Table~\ref{table_BFvars}. We use those variants to fix various parameters of the busy function. Note that we use a slight reformulation of the busy function, because we found it to be faster to fit with the LMA. We use the AIC to penalise each busy function variant's $\chi^{2}$ value for the number of free parameters and then choose the model with the best resultant $\chi^{2}$.
  
  The LMA was used to carry out the $\chi^{2}$ minimization for three reasons. Firstly, the LMA produces a $\chi^{2}$ covariance matrix that can be used to calculate both the uncertainties and correlations of the model parameters. Secondly, the LMA does not rely upon a grid of model parameter values, and the model parameters therefore do not suffer discretisation (such as in brute-force $\chi^{2}$ minimization). Finally, the LMA uses the $\chi^{2}$ covariance matrix to ensure that it simultaneously adjusts the model parameter values in a way that achieves the maximum decrement in $\chi^{2}$. As model dimensionality increases, this aspect of the LMA becomes increasingly powerful.
  
  We used variable re-mapping, singular value decomposition, and a modified power law to fit the busy function with the LMA. Variable re-mapping is necessary because the LMA cannot impose any range limits on model parameters. We used the variable re-mappings in Table~\ref{table_BFmaps} to limit the model parameters to sensible ranges and to avoid unphysical parameter values. This includes positive definite values for scaling/normalisation parameters and a user-specified range for the error function and power-law origins. We used singular value decomposition to ensure that degenerate model parameter values and extreme noise do not cause the LMA to fail.
  
  Using the LMA does not solve the problem common to all model fitting methods based on $\chi^{2}$ minimization: there is no guarantee that the LMA will find the global $\chi^{2}$ minimum when starting from an arbitrary position in parameter space. We solve this problem by using the LMA to find the nearest $\chi^{2}$ minimum for 1000~randomly chosen starting positions. These LMA starting positions are created in two steps. First, $\alpha$ is fixed at $10^{-3}$ and every other parameter is randomly chosen (with replacement) from the values in Table~\ref{table_LVMseeds}. The corresponding curve is then used to calculate an $\alpha$ and adjusted $\phi$ that ensure that the model peak is equal to a randomly chosen (with replacement) multiple of the peak data value (also listed in Table~\ref{table_LVMseeds}). We then assume that the model fit values of the smallest of these resultant $\chi^{2}$ values is a good proxy for the model fit values of the global $\chi^{2}$ minimum. An additional benefit of this approach is that we do not have to use too many iterations for each LMA process. The number of LMA iterations and starting positions can be traded against each other. This is because as the number of LMA starting positions is increased, the probability of choosing an LMA starting position close to the global $\chi^{2}$ minimum increases. For this application we are using 30~iterations for each LMA process compared to the $\mathrm{O}(1000)$ iterations typically used.
  
   To avoid local $\chi^{2}$ minima we have also added an additional criterion to the typical definition of a $\chi^{2}$ minimum. A $\chi^{2}$ minimum is typically defined as a negligible decrease in $\chi^{2}$ for consecutive iterations. In addition to this, we require that the $\chi^{2}$ value must not have increased for five consecutive iterations. This avoids $\chi^{2}$ minima that are `noise troughs' in unstable regions of parameter space.
   
   It should be noted that there is also a powerful, inherent advantage to model fitting via $\chi^{2}$ minimization. Model fitting with $\chi^{2}$ minimization takes into account the uncertainty of each individual data point. Our implementation includes this capability, although it was not required for fitting the HIPASS BGC spectra (which we assume to have a constant noise level of $13~\mathrm{mJy}$). We expect that this feature will be useful for datasets with channel/frequency-dependent noise. Alternatively, this capability can also be exploited by assigning large uncertainties to channels/frequencies affected by radio frequency interference. This will down-weight the significance of such channels/frequencies when parametrizing affected galaxies.
     
   Our implementation of the busy function fitting algorithm can be obtained from the project's website\footnote{http://code.google.com/p/busy-function-fitting/} or by sending an e-mail to one of the authors, R.~Jurek (russell.jurek@gmail.com). The implementation is available as a C~library, a C++~template library, and a Python module, all of which use \textsc{OpenMP} to take advantage of systems with multi-core CPUs.

    \section{Determination of uncertainties}
    \label{app_uncertainties}
    
    \begin{figure*}
        \centering
        \includegraphics[width=\linewidth]{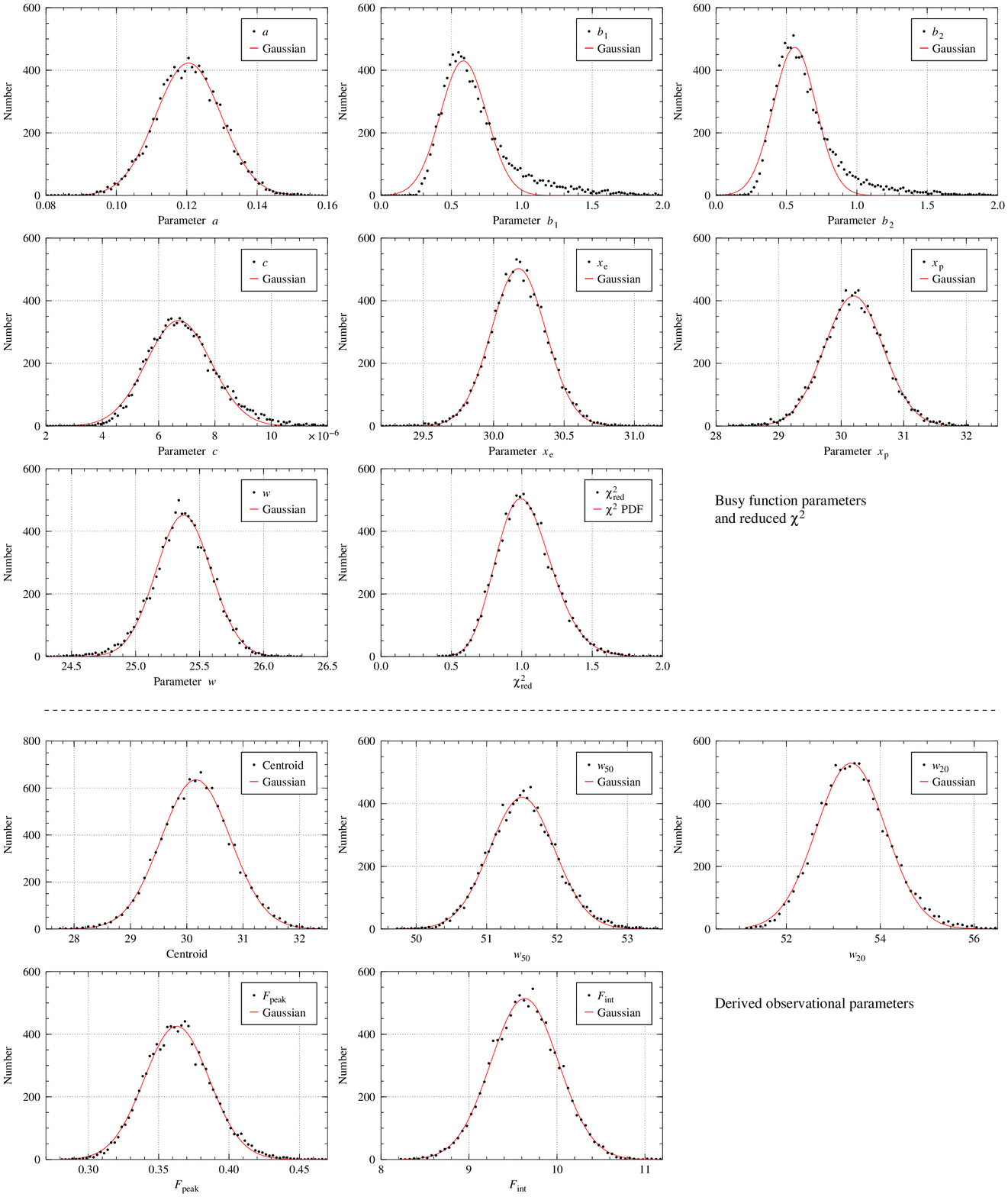}
        \caption{Distribution of busy function parameters and derived observational parameters after fitting the generalised busy function (with $n = 4$ fixed) to 10\,000 realisations of the integrated spectrum of NGC~3351 with artificial noise added. The red curve shows the result of a Gaussian fit to the parameter distribution, with the exception of the $\chi_{\rm red}^{2}$ distribution where we fitted the $\chi^{2}$~probability density function instead of a Gaussian. The unit of flux density is Jy, while spectral parameters are specified in channels.}
        \label{fig_parameters}
    \end{figure*}
    
    \begin{table*}
        \centering
        \caption{Busy function parameters (upper section) and derived observational parameters (lower section) after fitting the generalised busy function (with $n = 4$ fixed) to 10\,000 realisations of the integrated spectrum of NGC~3351 with artificial noise added. The first two columns list the parameter name and its value derived from fitting the original, high signal-to-noise spectrum. The uncertainties in the upper section were taken from the covariance matrix of the fit, while those in the lower section have been calculated with the new method introduced in Appendix~\ref{app_cholesky}. Columns~3 and~4 show the mean, $\bar{P}$, and median, $\tilde{P}$, of all 10\,000 realisations, columns~5 and~6 list the centroid, $P_{0}$, and standard deviation, $\sigma_{P}$, of a Gaussian function fitted to the parameter distribution (see Fig.~\ref{fig_parameters}), and column~7 shows the skewness, $\gamma_{1}$, of the distribution. The unit of flux density is Jy, while spectral parameters are specified in channels.}
        \label{tab_parameters}
        \begin{tabular}{lrrrrrr}
            \hline
            Parameter                & Original value     & $\bar{P}$ & $\tilde{P}$ & $P_{0}$  & $\sigma_{P}$ & $\gamma_{1}$ \\
            \hline
            $\varampl$               &  $0.121 \pm 0.002$ &  $0.121$  &  $0.121$    &  $0.120$ & $0.009$      &  $0.00$ \\
            $\varslop_{1}$           &  $0.604 \pm 0.036$ &  $0.705$  &  $0.617$    &  $0.585$ & $0.164$      &  $7.63$ \\
            $\varslop_{2}$           &  $0.572 \pm 0.034$ &  $0.651$  &  $0.585$    &  $0.558$ & $0.152$      &  $9.32$ \\
            $\varpoly \times 10^{6}$ &  $6.660 \pm 0.235$ &  $6.860$  &  $6.760$    &  $6.711$ & $1.151$      &  $0.63$ \\
            $x_{\rm e}$              & $30.183 \pm 0.040$ & $30.178$  & $30.176$    & $30.176$ & $0.194$      & $-0.14$ \\
            $x_{\rm p}$              & $30.206 \pm 0.098$ & $30.198$  & $30.200$    & $30.200$ & $0.470$      & $-0.02$ \\
            $w$                      & $25.387 \pm 0.045$ & $25.365$  & $25.369$    & $25.375$ & $0.215$      & $-0.39$ \\
            \hline
            Centroid                 & $30.164 \pm 0.122$ & $30.164$  & $30.168$    & $30.161$ & $0.614$      &  $0.00$ \\
            $w_{50}$                 & $51.612 \pm 0.085$ & $51.515$  & $51.510$    & $51.501$ & $0.462$      &  $0.17$ \\
            $w_{20}$                 & $53.485 \pm 0.130$ & $53.432$  & $53.399$    & $53.377$ & $0.736$      &  $0.32$ \\
            $F_{\rm peak}$           &  $0.348 \pm 0.005$ &  $0.364$  &  $0.363$    &  $0.363$ & $0.023$      &  $0.25$ \\
            $F_{\rm int}$            &  $9.618 \pm 0.075$ &  $9.633$  &  $9.634$    &  $9.632$ & $0.382$      &  $0.00$ \\
            \hline
        \end{tabular}
    \end{table*}
    
    \changed{A crucial aspect in the parametrization of spectral profiles of galaxies is the determination of uncertainties. Several different methods have been used in the parametrization of} \ion{H}{i}~\changed{lines in the past, including Monte-Carlo methods (e.g.~\citealt{Donley2005}) and bootstrap or jackknife methods (e.g.~\citealt{Hong2013}). These methods generally attempt to emulate repeated measurements by either adding artificial noise to the data or by looking at different subsets of a data set. While providing the most accurate assessment of uncertainties (apart from actually repeating the measurement), Monte-Carlo and bootstrap methods can be relatively slow due to the need to repeatedly alter the original data. In addition, not all data sets lend themselves to this type of procedure.}
    
    \changed{In this section we introduce two different methods that rely entirely on the covariance matrix of the busy function fit and can therefore be applied in situations where Monte-Carlo and bootstrap methods will fail, e.g.\ when only a single spectrum of low signal-to-noise ratio is available. In addition, the introduced methods are much faster than the former because they do not require the original data to be altered in any way.}
    
    \subsection{The classical approach: Monte-Carlo and bootstrap methods}
    \label{app_mc-bs}
    
    \changed{As Monte-Carlo and bootstrap methods} operate on the input data themselves, they are independent of the actual parametrization method used and can therefore be applied in combination with busy function fitting as well. As a demonstration of a Monte-Carlo approach, we created 10\,000~realisations of the integrated spectrum of the galaxy NGC~3351 \citep{Walter2008}, as shown in Fig.~\ref{fig_ngc300}, by adding random Gaussian noise with a standard deviation of $50~\mathrm{mJy}$ to the original THINGS spectrum. Next, we fitted the generalised busy function, $\varbgen$, with the parameter $n$ fixed to~4, to each of the 10\,000~spectra and extracted and analysed the resulting parameters. The results are summarised in Fig.~\ref{fig_parameters} and Table~\ref{tab_parameters}. The resulting $\chi^{2}$ of the fits follows the expected probability density function with a mean of $\langle \chi^{2} \rangle = 54.6$ and $\langle \chi_{\rm red}^{2} \rangle = 1.03$. The mean $\chi^{2}$ is slightly larger than the number of degrees of freedom of $53$ ($60$~spectral channels less $7$~free parameters), but this small discrepancy can be readily explained by intrinsic structure and noise in the original spectrum of NGC~3351.
    
    As shown in Fig.~\ref{fig_parameters}, most parameters obey an approximately normal distribution, although some are significantly skewed, in particular the two profile slope parameters, $\varslop_{1}$ and $\varslop_{2}$. Such non-Gaussian distributions imply that, strictly speaking, the parameter uncertainties cannot be expressed in terms of a single number, such as the standard deviation usually reported by least-squares fitting algorithms. However, as a first-order approximation, the parameters of the busy function can be assumed to follow a normal distribution. The same appears to be true for numerically derived observational parameters (line centroid, $w_{50}$ and $w_{20}$~line widths, peak flux density, and integrated flux) as presented in the lower sections of Fig.~\ref{fig_parameters} and Table~\ref{tab_parameters}.
    
    \subsection{A different approach: variation of busy function parameters}
    \label{app_cholesky}
    
    While Monte-Carlo and bootstrap methods provide a robust way of determining parameter uncertainties, it might not be possible to apply them in certain situations, e.g.~when only a single spectrum with low signal-to-noise ratio is available. In such situations, one of the great advantages of the busy function over direct \ion{H}{i}~parametrization methods comes into play: under the assumption of a Gaussian statistic of the busy function parameters we can determine the uncertainties of derived parameters, including line widths and fluxes, from the covariance matrix provided by least-squares fitting algorithms such as the Levenberg-Marquardt algorithm (LMA; \citealt{Levenberg1944,Marquardt1963}).
    
    As concluded in Appendix~\ref{app_analytical}, the lack of analytical solutions for derived parameters implies that we need to determine the uncertainties of derived parameters numerically. The simplest approach would be to randomly vary the busy function parameters coming out of a fit and recalculate derived parameters such as $w_{50}$ or $F_{\rm int}$. This can be repeated many times, and the uncertainties of the derived parameters can then be determined by either simply taking the standard deviation across all iterations or by fitting a Gaussian to the resulting parameter distribution. This approach implicitly assumes that all busy function parameters are normally distributed (see Appendix~\ref{app_mc-bs}) and that the individual parameters are entirely uncorrelated.
    
    Most parameters, however, have some degree of correlation (as shown in Fig.~\ref{fig_correlation} for the integrated spectrum of NGC~3351), and it is necessary to take this effect into consideration when randomly varying the fit parameters. This can be achieved with the following method based on the parameter covariance matrix provided by the LMA:
   
    \begin{figure}
        \centering
        \includegraphics[width=0.8\linewidth]{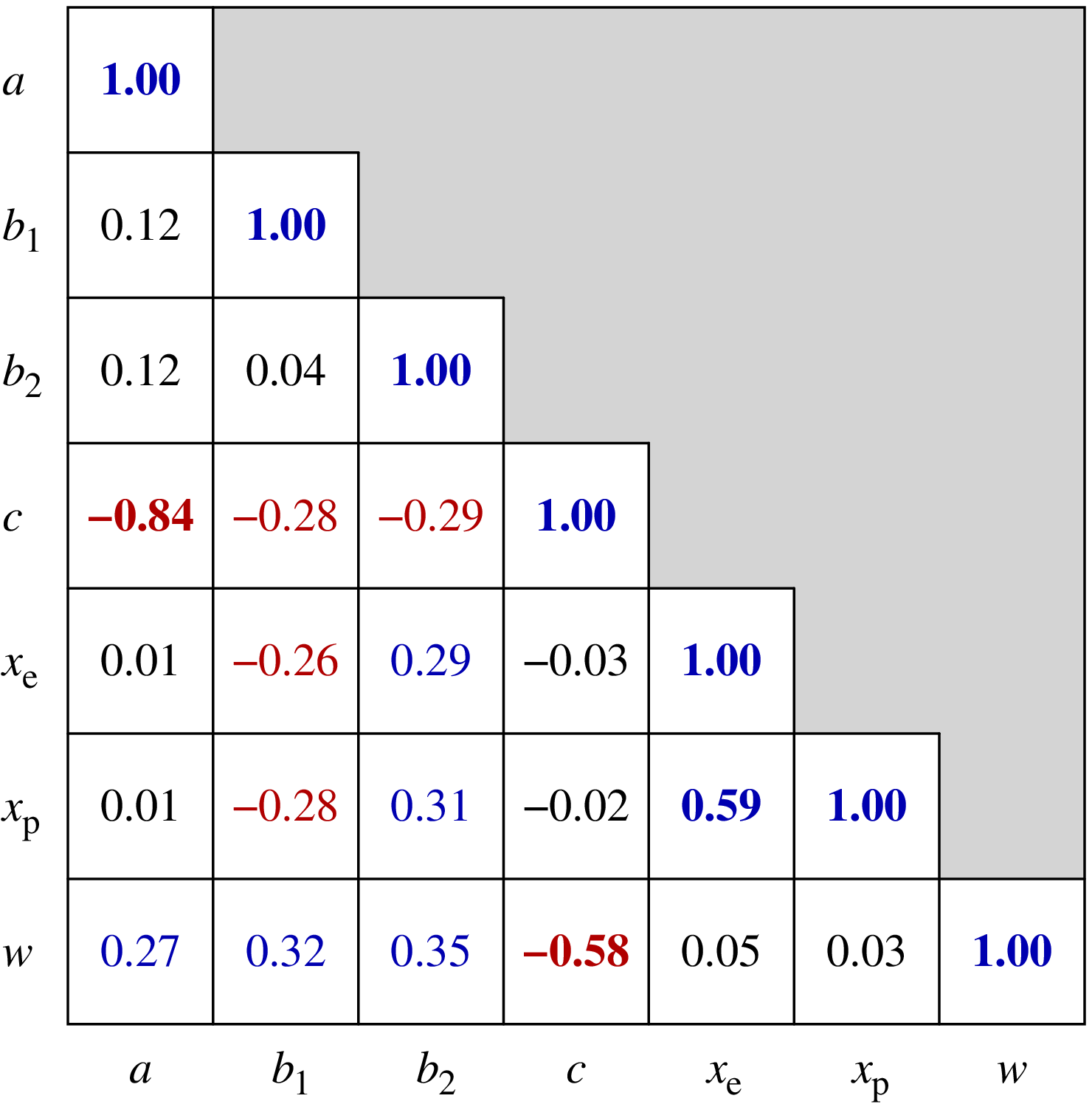}
        \caption{Correlation coefficients of generalised busy function parameters as derived from the covariance matrix of a fit to the integrated spectrum of NGC~3351.}
        \label{fig_correlation}
    \end{figure}
    
    \begin{enumerate}
        \item Using the LMA, the busy function is fitted to the integrated spectrum of the galaxy to be parametrized, providing us with values for the free parameters of the busy function, $\vec{p}$, as well as the full covariance matrix, $\matr{C}_{\vec{p}}$.
        \item Using the method of \citet{Box1958}, we can then generate $M$~sets of independent, random busy function parameters, $\vec{p}_{m}$, each following a normal distribution centred on $\vec{p}$ with a standard deviation as derived from the square root of the diagonal elements of $\matr{C}_{\vec{p}}$.
        \item Next, we need to transform the vectors of random, uncorrelated parameters, $\vec{p}_{m}$, into vectors of random parameters with the correct correlations, $\vec{p}_{m}'$, as described by the non-diagonal elements of $\matr{C}_{\vec{p}}$. This can be achieved by performing a Cholesky decomposition of the \textit{parameter correlation matrix},
        \begin{equation}
            \matr{K}_{\vec{p}} = \matr{L} \matr{L}^{\rm T} ,
        \end{equation}
        where $K_{\vec{p}}^{ij} = C_{\vec{p}}^{ij} / (\sigma_{i} \sigma_{j})$ and $\sigma_{i}^{2} = C_{\vec{p}}^{ii}$, and then multiplying each parameter vector with the \textit{lower triangular form} of the resulting matrix,
        \begin{equation}
            \vec{p}_{m}' = \matr{L} \vec{p}_{m} .
        \end{equation}
        In this step we need to take into account that each parameter in $\vec{p}_{m}$ must be of zero mean and unit variance for the transform to work. Hence, each parameter may need to be scaled and translated before and after the transform.
        \item Lastly, we can numerically derive the desired observational parameters (including centroid, $w_{50}$, $w_{20}$, $F_{\rm peak}$, and $F_{\rm int}$) for each of the $M$~correlated busy function parameter sets, $\vec{p}_{m}'$. We can then calculate the mean and standard deviation for each parameter across all $M$~iterations.
    \end{enumerate}
    
    We tested this parameter variation method on the integrated spectrum of NGC~3351 (Fig.~\ref{fig_ngc300}), fitting the generalised busy function, $\varbgen$, with the parameter $n = 4$ fixed. The results, listed in the lower section of Table~\ref{tab_parameters}, are in good agreement with our expectations, suggesting that the method produces accurate uncertainties. For example, when numerically measuring the flux density at the position of the central trough of the fitted busy function, we derive a value of $0.1210 \pm 0.001839$. Both the value and the uncertainty are identical (within the numerical accuracy) with those of the busy function parameter~$\varampl$ as derived from the least-squares fit. A similar comparison can be made between the measured line width, $w_{50} / 2 = 25.81 \pm 0.04226$, and the parameter $w = 25.39 \pm 0.04501$ of the busy function. Again, the values and uncertainties agree very well, even though we do not expect an exact identity of $w_{50} / 2$ and $w$ (see Appendix~\ref{app_analytical}). The results suggest that the error analysis method based on varying the initial busy function parameters produces accurate measurements of the uncertainty of derived parameters.
    
    \subsection{A faster approach: linear propagation of the covariance matrix}
    \label{app_linprop}
    
    \begin{table}
        \centering
        \caption{Comparison of the uncertainties of the observational parameters of the galaxy NGC~3351 as determined by the methods of parameter variation (Appendix~\ref{app_cholesky}) and error propagation (Appendix~\ref{app_linprop}). The last column lists the relative difference between the two methods. Flux parameters are specified in Jy, spectral parameters in channels.}
        \label{tab_uncertainties}
        \begin{tabular}{lrllr}
            \hline
            Parameter      & Value    & Uncert.    & Uncert.    & Difference \\
                           &          & par.~var.  & err. prop. & (per cent) \\
            \hline
            Centroid       & $30.164$ & $0.1216$   & $0.1213$   & $-0.25$    \\
            $w_{50}$       & $51.612$ & $0.08452$  & $0.08648$  & $ 2.32$    \\
            $w_{20}$       & $53.485$ & $0.1303$   & $0.1245$   & $-4.45$    \\
            $F_{\rm peak}$ &  $0.348$ & $0.004646$ & $0.005157$ & $11.00$    \\
            $F_{\rm int}$  &  $9.618$ & $0.07491$  & $0.07485$  & $-0.08$    \\

            \hline
        \end{tabular}
    \end{table}
    
    While the parameter variation method presented in Appendix~\ref{app_cholesky} provides an accurate way of estimating uncertainties, it is relatively slow and inefficient due to the large number of iterations required to achieve sufficient numerical accuracy. However, under the assumption of a \textit{linear approximation} of the function that translates between the free parameters of the busy function and the derived observational parameters of the spectral profile, we can instead use the error propagation law to determine not just the uncertainties of the derived parameters, but in fact the full covariance matrix.
    
    Let us assume that $\vec{p}$ is the parameter vector and $\matr{C}_{\vec{p}}$ the covariance matrix of the busy function's free parameters, as before. Let us further assume that there is a differentiable function, $\vec{f}$, that translates between the busy function parameters, $\vec{p}$, and the derived observational parameters, $\vec{q} = \vec{f}(\vec{p})$. We can then numerically approximate the Jacobian matrix of $\vec{f}$ by varying each input parameter, $p_{i}$, by a small amount, $\varepsilon_{i}$, such that
    \begin{equation}
        J_{ji} \equiv \frac{\partial f_{j}}{\partial p_{i}} \approx \frac{f_{j}(\vec{p} + \varepsilon_{i} \hat{\vec{e}}_{i}) - f_{j}(\vec{p})}{\varepsilon_{i}} ,
    \end{equation}
    where $\hat{\vec{e}}_{i}$ is the unit vector in the direction of the $i^{\text{th}}$ component of $\vec{p}$. With the Jacobian matrix determined, we can now use the error propagation law to calculate the full covariance matrix, $\matr{C}_{\vec{q}}$, of the derived observational parameters:
    \begin{equation}
        \matr{C}_{\vec{q}} = \matr{J} \matr{C}_{\vec{p}} \matr{J}^{\rm T} .
    \end{equation}
    The uncertainties of the individual parameters, $q_{i}$, are then given by the square root of the diagonal elements of the covariance matrix. Note that, for this method to work, it is not necessary to know the analytic expression of $\vec{f}(\vec{p})$, as the function can be evaluated numerically.
    
    As before, we tested the method of propagating the covariance matrix on the integrated spectrum of NGC~3351 (Fig.~\ref{fig_ngc300}), using relative offsets of $\varepsilon_{i} = |p_{i}| \times 10^{-5}$. The results, including a comparison with the method introduced in Appendix~\ref{app_cholesky}, are presented in Table~\ref{tab_uncertainties}. Both the method of parameter variation as well as the method of linear propagation of the covariance matrix yield comparable estimates of the uncertainties of the derived observational parameters of NGC~3351. The largest discrepancy is observed for the peak flux density, $F_{\rm peak}$, for which the error propagation method yields an uncertainty that is by about $10$~per cent higher than that of the parameter variation method. These small discrepancies are likely due to the linear approximation made in the error propagation method.
    
\begin{figure}
        \centering
        \includegraphics[width=0.6\linewidth]{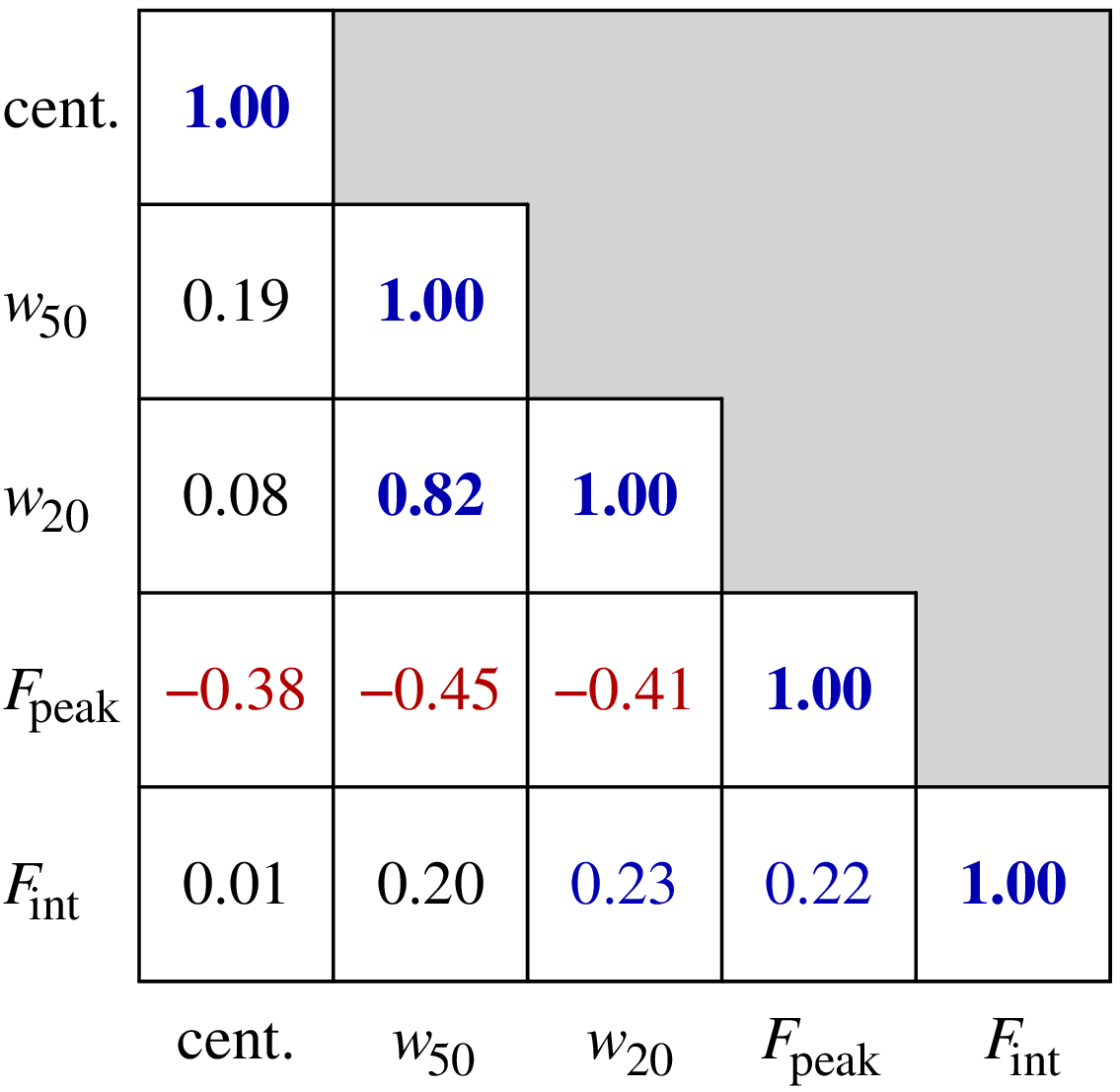}
        \caption{Correlation coefficients of observational parameters derived from a linear propagation of the covariance matrix resulting from a fit to the integrated spectrum of NGC~3351.}
        \label{fig_correlation_2}
    \end{figure}

    The error propagation method is generally much faster than the parameter variation method described in Appendix~\ref{app_cholesky}, as it does not require a large number of numerical iterations. Another advantage of the error propagation method is that it will produce a full parameter covariance matrix `for free' (i.e.\ without the need for computationally expensive numerical iterations), thus providing information about correlations between observational parameters. This is illustrated in Fig.~\ref{fig_correlation_2}, where the correlation coefficients derived from the covariance matrix, $\matr{C}_{\vec{q}}$, of the fit to the spectrum of NGC~3351 are presented.
    
    \subsection{Conclusions}
    
    The possibility of deriving uncertainties from the covariance matrix of a busy function fit, either by randomly varying the initial fit parameters or by linearly propagating the covariance matrix, turns the busy function into a powerful tool for measuring the observational parameters of galaxies from their integrated spectra. The great advantage of the presented methods is that, unlike classical Monte-Carlo or bootstrap methods, they \changed{are computationally inexpensive and} do not require alterations to the data being fitted. Hence, both methods can be applied in situations where the input data are not suitable for classical error analysis methods, e.g.~when only a single spectrum of low signal-to-noise ratio is available.
    
    There are several important limitations to both methods, including the implicit assumption of Gaussian errors for all parameters as well as the requirement to have a sufficiently large number of samples across the width of the profile, as otherwise the fit will be under-determined, resulting in unrealistic uncertainty estimates. These limitations, however, are not specific to the busy function, but generally apply whenever analytic functions are fitted to data. In the case of the error propagation method we need to make the additional assumption of a linear propagation of uncertainties. This assumption will break down in certain situations where highly non-linear relations between parameters exist, potentially leading to unrealistic uncertainty estimates, in particular when the uncertainties of the fitted busy function parameters are large compared to their values. Our tests of the two methods on the integrated spectrum of NGC~3351 demonstrate that both yield accurate uncertainty estimates when applied to well-resolved spectral profiles. However, it is important to keep the limitations discussed above in mind when using either method in the determination of parameter uncertainties.
    
    \bsp
    \label{lastpage}
\end{document}